\documentclass[10pt,letterpaper]{article}
\usepackage[top=0.85in,left=2.75in,footskip=0.75in,marginparwidth=2in]{geometry}

\usepackage[utf8]{inputenc}
\usepackage{cite}

\usepackage{nameref,hyperref}

\usepackage{microtype}
\DisableLigatures[f]{encoding = *, family = * }

\raggedright
\usepackage{setspace} 
\usepackage{changepage}

\usepackage[aboveskip=1pt,labelfont=bf,labelsep=period,singlelinecheck=off]{caption}

\makeatletter
\renewcommand{\@biblabel}[1]{\quad#1.}
\makeatother

\usepackage{lastpage,fancyhdr,graphicx}
\usepackage{epstopdf}
\fancyheadoffset[L]{2.25in}
\fancyfootoffset[L]{2.25in}

\usepackage{color}

\definecolor{Gray}{gray}{.25}

\usepackage{graphicx}

\usepackage{sidecap}

\usepackage[round]{natbib}
\usepackage{amsmath, amsthm, amssymb, amsfonts}
\usepackage{datetime}

\usepackage{wrapfig}
\usepackage[pscoord]{eso-pic}
\usepackage[fulladjust]{marginnote}
\reversemarginpar

\begin{document}
\vspace*{0.35in}

\begin{flushleft}
{\Large
\textbf\newline{Physical Basis of Large Microtubule Aster Growth}
}
\newline
\singlespacing
Keisuke Ishihara\textsuperscript{1, 2, 3, *},
Kirill S. Korolev\textsuperscript{4,*},
Timothy J. Mitchison\textsuperscript{1, 2} \\
\bigskip
\textsuperscript{1} Department of Systems Biology, Harvard Medical School \\
200 Longwood Avenue, Boston, MA, USA 02115 \\
\bigskip
\textsuperscript{2} Cell Division Group, Marine Biological Laboratory \\
7 MBL Street, Woods Hole, MA, USA 02543\\
\bigskip
\textsuperscript{3} Present address:
	Center for Systems Biology Dresden\\
	Pfotenhauerstr. 108, Dresden, Germany 01307\\
\bigskip
\textsuperscript{4} Department of Physics and Graduate Program in Bioinformatics\\
	Boston University, Boston, MA, USA 02215
\\
\bigskip
\textsuperscript{*} Corresponding authors:
ishihara@mpi-cbg.de, korolev@bu.edu
\\
\bigskip
version \today

\end{flushleft}

\doublespacing

\section*{Abstract}
Microtubule asters - radial arrays of microtubules organized by centrosomes - play a fundamental role in the spatial coordination of animal cells. The standard model of aster growth assumes a fixed number of microtubules originating from the centrosomes. However, aster morphology in this model does not scale with cell size, and we recently found evidence for non-centrosomal microtubule nucleation. Here, we combine autocatalytic nucleation and polymerization dynamics to develop a biophysical model of aster growth. Our model predicts that asters expand as traveling waves and recapitulates all major aspects of aster growth. As the nucleation rate increases, the model predicts an explosive transition from stationary to growing asters with a discontinuous jump of the growth velocity to a nonzero value.  Experiments in frog egg extract confirm the main theoretical predictions. Our results suggest that asters observed in large frog and amphibian eggs are a meshwork of short, unstable microtubules maintained by autocatalytic nucleation and provide a paradigm for the assembly of robust and evolvable polymer networks.


\section*{INTRODUCTION}

Animal cells use asters, radial arrays of microtubules, to spatially organize their cytoplasm \citep{Wilson:1896vp}. Specifically, astral microtubules transport organelles \citep{WatermanStorer:1998vw, Wang:2013ds, Grigoriev:2008vg}, support cell motility by mediating mechanical and biochemical signals \citep{EtienneManneville:2013gi}, and are required for proper positioning of  the nucleus, the mitotic spindle, and the cleavage furrow \citep{Field:2015fh, Grill:2005kv, Neumuller:2009ee, Wilson:1896vp, Tanimoto:2016db}. Within asters, individual microtubules undergo dynamic instability \citep{Mitchison:1984ub}: They either grow (polymerize) or shrink (depolymerize) at their plus ends and stochastically transition between these two states. Collective behavior of microtubules is less well understood, and it is not clear how dynamic instability of individual microtubules controls aster growth and function. 

The standard model of aster growth posits that centrosomes nucleate and anchor all microtubules at their minus ends while the plus ends polymerize outward via dynamic instability \citep{Brinkley:1985il}. As a result, aster growth is completely determined by the dynamics of individual microtubules averaged over the growing and shrinking phases. In particular, the aster either expands at a velocity given by the net growth rate of microtubules or remains stationary if microtubules are unstable and tend to depolymerize \citep{Dogterom:1993tz, Verde:1992tl, Belmont:1990vs}. 

The standard model of aster growth is being increasingly challenged by reports of microtubules with their minus ends located far away from centrosomes \citep{Keating:1999tja, Akhmanova:2015ct}. Some of these microtubules may arise simply by detachment from centrosomes \citep{Keating:1997vb, WatermanStorer:2000to} or severing of pre-existing microtubules \citep{RollMecak:2010kx}. However, new microtubules could also arise due to a nucleation processes independent of centrosomes \citep{Efimov:2007hg, Clausen:2007fi, Petry:2013iu} and contribute to both aster growth and its mechanical properties. Indeed, we recently observed that centrosomal nucleation is insufficient to explain the large number of growing plus ends found in asters \citep{Ishihara:2014ih}. Moreover, the standard model demands a decrease in microtubule density at aster periphery, which is inconsistent with aster morphology in frog and fish embryos \citep{Wuhr:2008dz, Wuhr:2010fi}. To resolve these inconsistencies, we proposed an autocatalytic nucleation model, where microtubules or microtubule plus ends stimulate the nucleation of new microtubules at the aster periphery \citep{Wuhr:2009wv, Ishihara:2014ih, Ishihara:2014gpa}. This mechanism generates new microtubules necessary to maintain a constant density as the aster expands. We also hypothesized that autocatalytic nucleation could effectively overcome extinction of individual microtubules, and allow rapid growth of large asters made of short, unstable microtubules. However, we did not provide a quantitative model that can be compared to the experiments or even show that the proposed mechanism is feasible.

Here, we develop a quantitative biophysical model of aster growth with autocatalytic nucleation. It predicts that asters can indeed expand even when individual microtubules turn over and disappear by depolymerization. In this regime, aster expansion is driven by the increase in the total number of microtubules, and the resulting aster is a network of short interconnected microtubules. The transition from stationary to growing asters depends on the balance between polymerization dynamics and nucleation. At this transition, our theory predicts a minimum rate at which asters grow, which we define as the gap velocity. This gap velocity arises due to the dynamic instability of microtubule polymerization and excludes a wide class of alternative models. More importantly, this mode of aster growth allows the cell to assemble asters with varying polymer densities at consistently large speeds. Using a cell-free reconstitution approach \citep{Nguyen:2014fi, Field:2014eb}, we perform biochemical perturbations and observe the slowing down and eventual arrest of aster growth with a substantial gap velocity at the transition. By combining theory and experiments, we provide a quantitative framework for how the cell cycle may regulate the balance between polymerization dynamics and nucleation to control aster growth. We propose that the growth of large interphase asters is an emergent property of short microtubules that constantly turnover and self-amplify.

\section*{RESULTS}

\subsection*{Conceptual Model for Aster Growth based on Polymerization Dynamics and Autocatalytic Nucleation}

Asters are large structures comprised of thousands of microtubules. How do the microscopic dynamics of individual microtubules determine the collective properties of asters such as their morphology and growth rate? Can asters sustain growth when individual microtubules are unstable? To address these questions, we develop a theoretical framework that integrates polymerization dynamics and autocatalytic nucleation (Fig. 1A). Our main goal is to determine the distribution of microtubules within asters and the velocity at which asters grow:

\begin{equation}
V=\dfrac{d\,Radius}{dt}.
\label{eq:VeqdRdt}
\end{equation}

Beyond being the main experimental readout, the aster growth velocity is crucial for cell physiology because it allows large egg cells to divide its cytoplasm rapidly.

Polymerization dynamics of plus ends is an individual property of microtubules. To describe plus end dynamics, we adopt the two-state model of microtubule dynamic instability (Fig. 1A, left). In this model, a single microtubule is in one of the two states: (i) the growing state, where plus ends polymerize at rate $v_{g}$ and (ii) the shrinking state, where plus ends depolymerize at rate $v_{s}$. A growing microtubule may transition to a shrinking state (catastrophe event) with rate $f_{cat}$. Similarly, the shrinking to growing transition (rescue event) occurs at rate $f_{res}$. For large asters growing in \textit{Xenopus} egg cytoplasm, we provide estimates of these parameters in Table 1.

Plus end dynamics can be conveniently summarized by the time-weighted average of the polymerization and depolymerization rates \citep{Verde:1992tl, Dogterom:1993tz}:

\begin{equation}
J=\frac{v_{g}f_{res}-v_{s}f_{cat}}{f_{res}+f_{cat}}.
\label{eq:defineJ}
\end{equation}

This parameter describes the tendency of microtubules to grow or shrink. When $J<0$, microtubules are said to be in the bounded regime because their length inevitably shrinks to zero, i.e. microtubule disappears. When $J>0$, microtubules are said to be in the unbounded regime, because they have a nonzero probability to become infinitely long. Parameter $J$ also determines the mean elongation rate of a very long microtubule that persists over many cycles of catastrophe and rescue. The dynamics of short microtubules, however, depends on their length and initial state (growing vs. shrinking) and should be analyzed carefully.

The standard model posits that asters are produced by the expansion of individual microtubules, so the transition from small mitotic asters to large interphase asters is driven by a change in the sign of $J$ \citep{Verde:1992tl, Dogterom:1993tz} (Fig. 1B left, “individual growth”). With bounded dynamics $J<0$, the standard model predicts that every microtubule shrinks to zero length and disappears. This microtubule loss is balanced by nucleation of new microtubules at the centrosomes, the only place where nucleation is allowed in the standard model. As a result, asters remain in the stationary state and are composed of a few short microtubules, and the aster growth velocity is thus $V=0$. With unbounded dynamics $J>0$, the standard model predicts an aster that has a constant number of microtubules and increases its radius at a rate equal to the elongation rate of microtubules (i.e. $V=J$).

Below, we add autocatalytic microtubule nucleation (Fig. 1A, right) to the standard model and propose the regime of “collective growth” (Fig. 1B, right). In this regime, asters grow ($V>0$) although individual microtubules are bounded ($J<0$) and are, therefore, destined to shrink and disappear. The growth occurs because more microtubules are nucleated than lost, and new microtubules are typically nucleated further along the expansion direction. Indeed, when a new microtubule is nucleated, it is in a growing state and starts expanding outward before its inevitable collapse. During its lifetime, this microtubule can nucleate a few more microtubules all of which are located further along the expansion direction. As we show below, this self-amplifying propagation of microtubules is possible only for sufficiently high nucleation rates necessary to overcome microtubule loss and sustain collective growth.

Specifically, we assume that new microtubules nucleate at locations away from centrosomes at rate $Q$. This rate could depend on the local density of growing plus ends if they serve as nucleation sites or the local polymer density if nucleation can occur throughout a microtubule.  This rate could depend on the local density of growing plus ends, if they serve as nucleation sites or the local polymer density if nucleation occurs along the side of pre-existing microtubules. The new microtubules have zero length and tend to grow radially due to mechanical interactions with the existing microtubule network. These non-centrosomal microtubules disappear when they shrink back to their minus ends. Our assumptions are broadly consistent with known microtubule physiology \citep{Clausen:2007fi, Petry:2013iu}, and we found strong evidence for nucleation away from centrosomes in egg extract by microtubule counting in growing asters \citep{Ishihara:2014ih}. Below, we consider plus-end-stimulated nucleation and the analysis for the polymer-stimulated nucleation is summarized in the SI.

\begin{figure}[ht]
\includegraphics[width=0.8\textwidth]{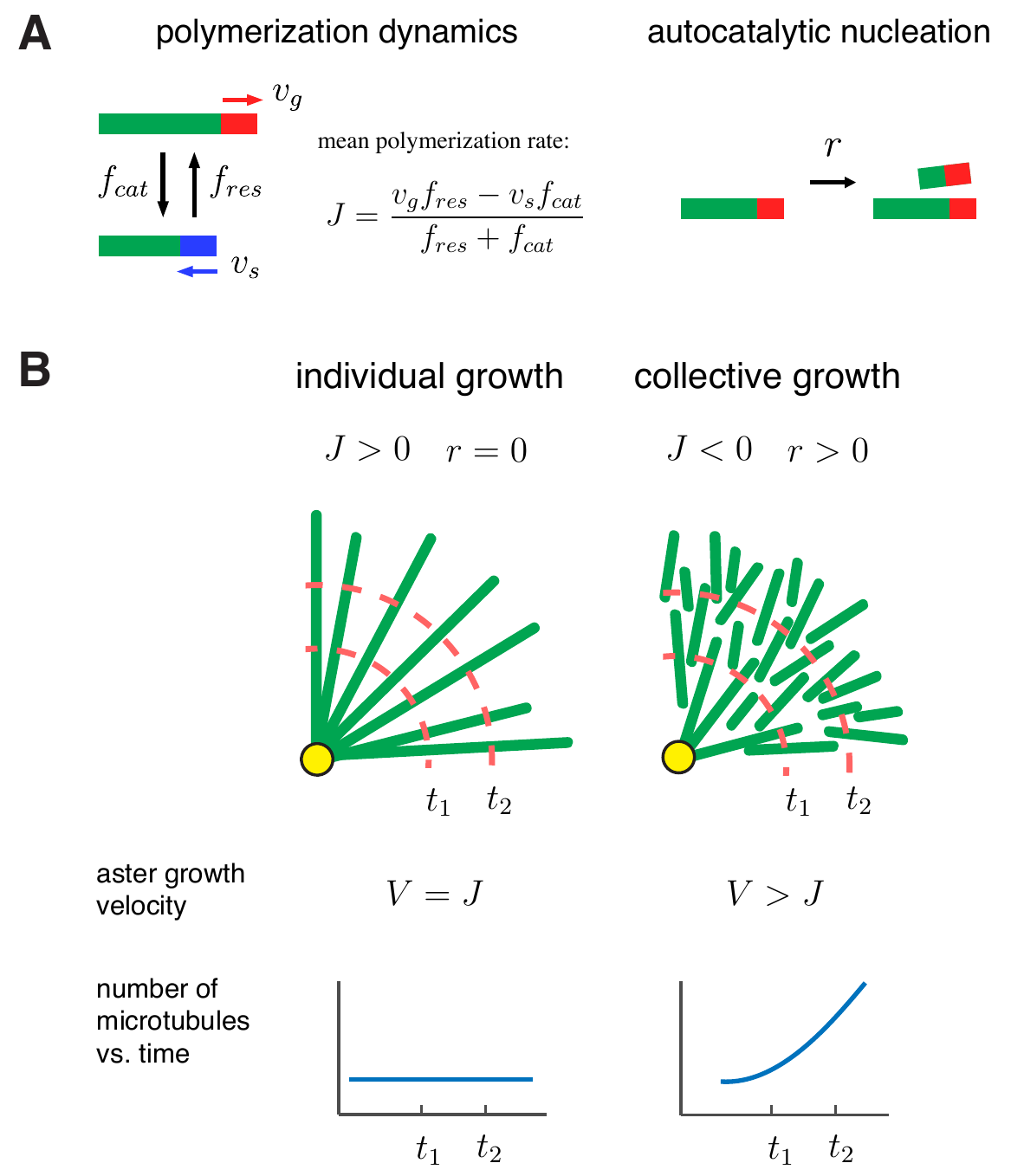}
\caption{\color{Gray} 
\textbf{A biophysical model for the collective growth of microtubule asters.}
(A) We propose that asters grow via two microscopic processes: polymerization and nucleation. Individual microtubules follow the standard dynamic instability with a growing sate with polymerization rate $v_{g}$ and a shrinking state with depolymerization rate $v_{s}$. Transitions between the states occur at rates $f_{cat}$ and $f_{res}$, which model catastrophe and rescue events, respectively. New microtubules are added at a rate $r$ via a nucleation at pre-existing plus ends in the growing state. (B) Individual vs. collective growth of asters. In the standard model of “individual growth”, asters increase their radius at rate $V=\dfrac{d\,Radius}{dt}$ only via a net polymerization from the centrosome (yellow). Thus, this model predicts that the rate of aster growth equals the mean polymerization rate $V=J$, the number of microtubules is constant, and their density decreases away from the centrosomes. In the collective growth model, the microtubule density is constant and the number of microtubules increases. Autocatalytic nucleation makes asters grow faster than the net polymerization rate $J$ and can sustain growth even when individual microtubules are unstable $J<0$.}
\label{cartoon}
\end{figure}

Without negative feedback, autocatalytic processes lead to exponential growth, but there are several lines of evidence for an apparent “carrying capacity” of microtubules in a given cytoplasmic volume \citep{Clausen:2007fi, Petry:2013iu, Ishihara:2014ih}. Saturation is inevitable since the building blocks of microtubules are present at a fixed concentration. In our model, we impose a carrying capacity by expressing autocatalytic nucleation as a logistic function of the local density of growing plus ends, which is qualitatively consistent with local depletion of nucleation factors such as the gamma-tubulin ring complex. Other forms of negative feedback (e.g. at the level of polymerization dynamics) are possible as well. In SI, we show that the type of negative feedback does not affect the rate of aster growth, which is determined entirely by the dynamics at the leading edge of a growing aster where the microtubule density is small and negative feedback can be neglected.

\subsection*{Mathematical Model of Autocatalytic Growth of Asters}

Assuming large number of microtubules, we focus on the mean-field or deterministic dynamics (SI) and formalize our model as a set of partial differential equations. Specifically, we let $\rho_{g}(t, x, l)$ and $\rho_{s}(t, x, l) $ denote respectively the number of growing and shrinking microtubules of length $l$ with their minus ends at distance $x>0$ from the centrosome. The number of newly nucleated microtubules is given by $Q(x)=rC_{g}(t,x)(1-C_{g}(t,x)/K)$, where $r$ is the nucleation rate, $K$ is the carrying capacity controlling the maximal plus end density, and $C_{g}(t,x)$ is the local density of the growing plus ends at point $x$. The polymerization dynamics and nucleation are then described by,
 
\begin{equation}
\left\{ \ \begin{aligned}\frac{\partial\rho_{g}}{\partial t}=-v_{g}\frac{\partial\rho_{g}}{\partial l}-f_{cat}\rho_{g}+f_{res}\rho_{s} & +Q(x)\cdot\delta(l)\\
\frac{\partial\rho_{s}}{\partial t}=+v_{s}\frac{\partial\rho_{s}}{\partial l}+f_{cat}\rho_{g}-f_{res}\rho_{s}.
\end{aligned}
\right.\label{eq:PDEmodel}
\end{equation}

Note that polymerization and depolymerization changes the microtubule length $l$, but not the minus end position $x$. Equations at different $x$ are nevertheless coupled due to the nucleation term, which depends on $x$ through $C_{g}$. 

To understand this system of equations, consider the limit of no nucleation ($r=0$). Then, the equations at different $x$ become independent and we recover the standard model that reduces aster growth to the growth of individual microtubules \citep{Verde:1992tl, Dogterom:1993tz}. With nucleation, aster growth is a collective phenomenon because microtubules of varying length and minus end positions contribute to $C_{g}(t,x)$, which can be expressed as a convolution of $\rho_{g}$ (see SI). The delta-function $\delta(l)$ ensures that newly nucleated microtubules have zero length. 

Finally, we need to specify what happens when microtubules shrink to zero length. In our model, microtubules originating from centrosomes rapidly switch from shrinking to growth (i.e. re-nucleate), while non-centrosomal microtubules disappears completely (i.e. no re-nucleation occurs). We further assume that mother and daughter microtubules disappear without affecting each other. Indeed, if the collapse of the mother microtubule triggered the collapse of the daughter microtubule (or vice versa), then no net increase in the number of microtubules would be possible in the bounded regime. One consequence of this assumption is that the minus end of a daughter microtubule becomes detached from any other microtubules in the aster following the collapse of the mother microtubule. As a result, minus ends need to be stabilized after nucleation possibly by some additional factors \citep{Akhmanova:2015iw} and mechanical integrity of the aster should rely on microtubule bundling \citep{Ishihara:2014ih}. 

\subsection*{Asters Can Grow as Spatially Propagating Waves with Constant Bulk Density}

To check if our model can describe aster growth, we solved Eq. \eqref{eq:PDEmodel} numerically using finite difference methods in an 1D planar geometry. With relatively low nucleation rates and $J<0$, microtubule populations reached a steady-state profile confined near the centrosome reminiscent of an aster in the standard model with bounded microtubule dynamics (Fig. 2A left). When the nucleation rate was increased, the microtubule populations expanded as a travelling wave with an approximately invariant shape and constant microtubule density at the periphery (Fig. 2A right) consistent with the growth of interphase asters in our reconstitution experiments (Fig. 2B and \citep{Ishihara:2014ih}). Thus, our model predicted two qualitatively different states: stationary and growing asters.

\begin{figure}[ht]
\includegraphics[width=\textwidth]{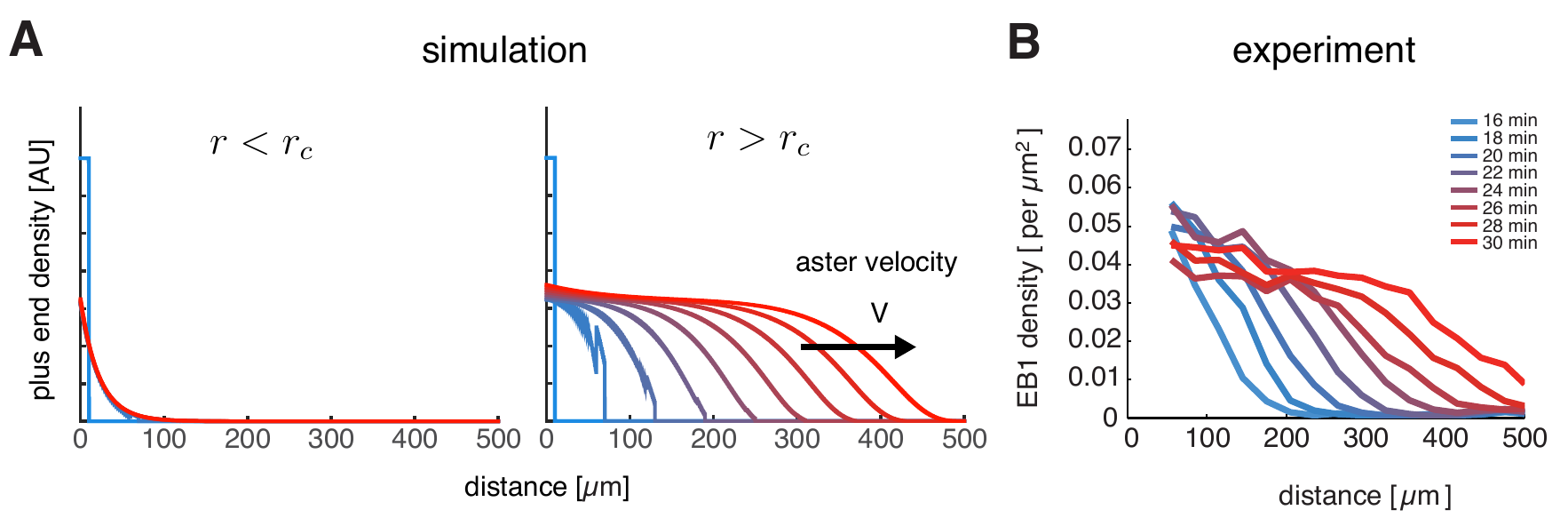}
\caption{\color{Gray}
\textbf{Our model captures key features of large aster growth.}
(A) Time evolution of growing plus end density predicted by our model, which we solved via numerical simulations in 1D geometry. In the stationary regime, the microtubule population remained near the centrosome $v_{g}=30$, $v_{s}=40$, $f_{cat}=3$, $f_{res}=1$, and $r=1.0$ (left). In contrast, outward expansion of the microtubule population was observed when the nucleation rate was increased to $r=2.5$, above the critical nucleation rate $r_{c}$ (right). For both simulations, microtubules are in the bounded regime $J<0$. (B) Experimental measurements confirm that asters expand at a constant rate with time-invariant profiles of the plus end density, as predicted by our model. The plus end densities were estimated as EB1 comet density during aster growth as previously described \citep{Ishihara:2014ih}. © 2014, Proceedings of the National Academy of Sciences of the United States of America. All Rights Reserved. Panel B reprinted with permission from Figure 4C from \citep{Ishihara:2014ih}, Proceedings of the National Academy of Sciences of the United States of America. Not covered by the terms of the Creative Commons Attribution 4.0 International license. 
}
\label{simuexpt}
\end{figure}

\subsection*{Analytical Solution for Growth Velocity and Critical Nucleation}

Next, we solved Eq. \eqref{eq:PDEmodel} exactly and obtained the following analytical expression for the growth velocity of an aster in terms of model parameters:

\begin{equation}
V=\frac{v_{g}(v_{g}f_{res}-v_{s}f_{cat})^{2}}{\left(\begin{aligned}v_{g}(v_{g}f_{res}-v_{s}f_{cat})(f_{res}+f_{cat})+(v_{g}+v_{s})(v_{g}f_{res}+v_{s}f_{cat})r\\
-2(v_{g}+v_{s})\sqrt{v_{g}f_{cat}f_{res}r(v_{g}f_{res}-v_{s}f_{cat}+v_{s}r)}
\end{aligned}
\right)},
\label{eq:velocity}
\end{equation}

which holds for the parameter range $r_{c}<r<f_{cat}$. The details of the calculation, including the definition of $r_{c}$ are summarized in SI.
\begin{figure}[ht]
\includegraphics[width=\textwidth]{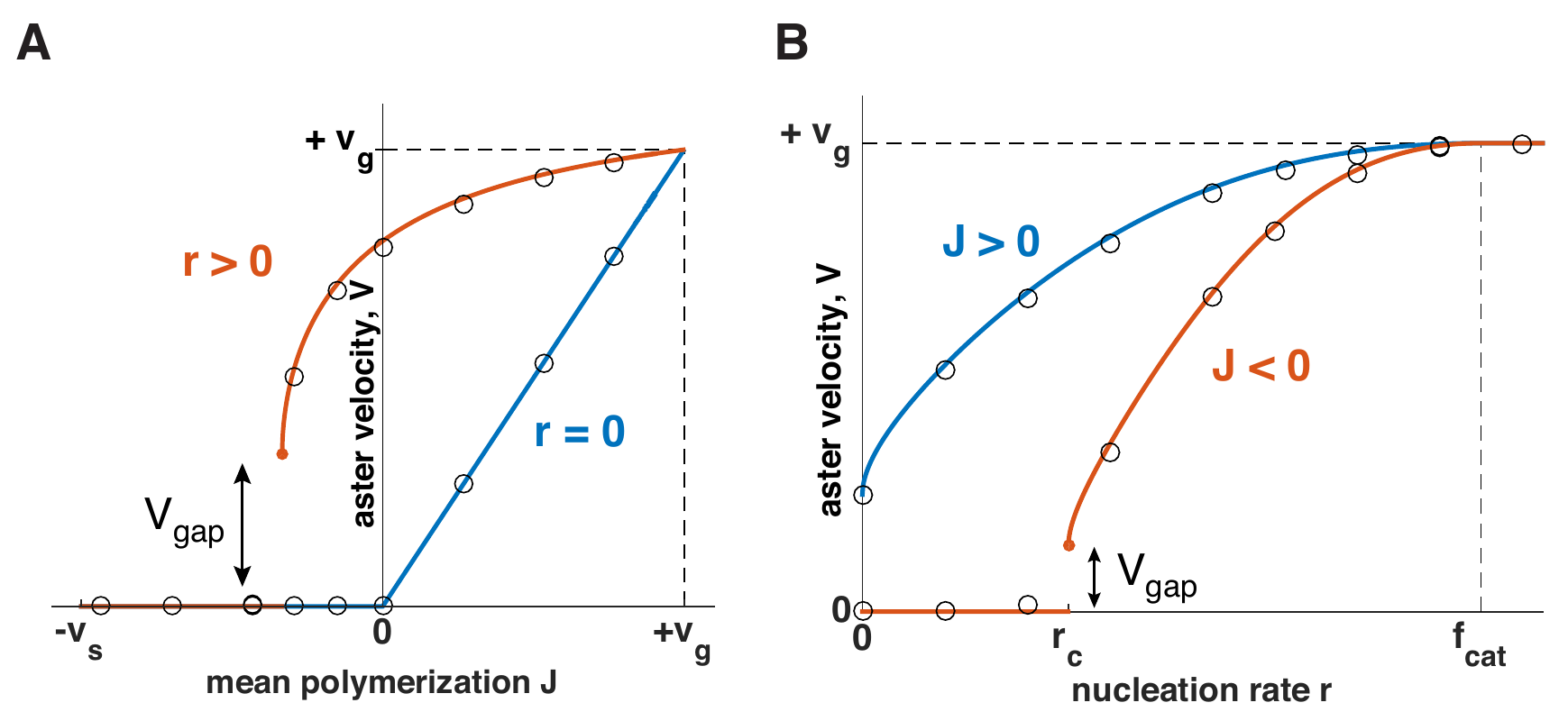}
\caption{\color{Gray} 
\textbf{Explosive transition from stationary to growing asters and other theoretical predictions.} Analytical solution (lines) and numerical simulations (dots) predict that asters either remain stationary or expand at a constant velocity, which increases with the net polymerization rate $J$ (A) and nucleation rate $r$ (B). The transition to a growing state is accompanied by finite jump in the expansion velocity labeled as $V_{gap}$. (A) The behavior in the standard model ($r = 0$) is shown in blue and our model ($r = 1.5$) in red. Note that aster growth commences at $J<0$ in the presence of nucleation and occurs at a minimal velocity $V_{gap}$. Although spatial growth can occur for both $J>0$ and $J<0$ the properties of the resulting asters could be very different (see SI). Here, $v_{g}=30, v_{s}=30, f_{cat} = 3$. (B) If $J<0$, critical nucleation $r_{c}$ is required to commence aster growth. Blue line corresponds to $J>0 (v_{g}=30, v_{s}=15, f_{cat} = 3, f_{res}=3)$ and red line to $J<0 (v_{g}=30, v_{s}=15, f_{cat} = 3, f_{res}=1)$. See Materials and Methods and SI for the details of the analytical solution and numerical simulations.
}
\label{VJVr}
\end{figure}

Using this expression, we summarize how aster growth velocity $V$ is affected by the mean polymerization rate $J$ (Fig. 3A) and nucleation rate $r$ (Fig. 3B). In the absence of autocatalytic nucleation ($r=0$), our model reduces to the standard model and predicts that asters only grow when $J>0$ with $V=J$ (Fig. 3A blue line). When nucleation is allowed ($r>0$), the growth velocity increases with $r$ and asters can grow even when individual microtubules are unstable $J<0$ (Fig. 3A and 3B). During this collective growth, the aster expands radially because more microtubules are nucleated than lost at the front. In the aster bulk, nucleation is reduced from the carrying capacity, and the aster exists in the dynamic balance between microtubule gain due to nucleation and loss due to depolymerization. Since microtubules are in the bounded regime, their lifetime is short, and they disappear before reaching appreciable length. In sharp contrast to the standard model, we predict that asters are a dynamic network of short microtubules with properties independent from the distance to the centrosome. Thus, nucleation not only increases the number of microtubules, but also controls the growth rate and spatial organization of asters enabling them to span length scales far exceeding the length of an individual microtubule. 

When $J<0$, a critical nucleation rate is required for aster growth (Fig. 3B). Indeed, microtubules constantly disappear as their length shrinks to zero, and the nucleation of new microtubules need to occur frequently enough to overcome the microtubule loss. Consistent with this argument, our analytical solution predicts no aster growth below a certain value of nucleation (SI), termed critical nucleation rate $r_{c}$:
 
\begin{equation}
r_{c}=f_{cat}-\dfrac{v_{g}}{v_{s}}f_{res}\textrm{.}\label{eq:rcritical}
\end{equation}
The right hand side of this equation is the inverse of the average time that a microtubule spends in the growing state before shrinking to zero-length and disappearing (SI). Thus, aster growth requires that, on average, a microtubule to nucleate at least one new microtubule during its lifetime.

The dependence of the critical nucleation rate on model parameters is very intuitive.  Increasing the parameters in favor of polymerization ($v_{g}$ and $f_{res}$), lowers the threshold level of nucleation required for aster growth, while increasing the parameters in favor of depolymerization ($v_{s}$ and $f_{cat}$) has the opposite effect. We also find that $r_{c}=0$ when J = 0, suggesting that there is no critical nucleation rate for $J\geq0$. This limit is consistent with the standard model with $J>0$ and $r=0$ where the aster radius increases albeit with radial dilution of microtubule density (Fig. 1B). The critical nucleation rate conveys the main implication of our theory: the balance between polymerization dynamics and autocatalytic nucleation defines the quantitative condition for continuous aster growth. 

\subsection*{Explosive Transition to Growth with a ``Gap Velocity"}
At the critical nucleation rate $r=r_c$, the aster growth velocity $V$ takes a positive, nonzero value (Fig. 3), which we refer to as the ``gap velocity" (SI): 
 
\begin{equation}
V_{gap}\equiv\underset{r\rightarrow r_{c}}{\lim}V=\dfrac{-v_{g}v_{s}(v_{g}f_{res}-v_{s}f_{cat})}{v{}_{g}^{2}f_{res}+v{}_{s}^{2}f_{cat}} \textrm{.}\label{eq:vgap}
\end{equation}

This finite jump in the aster velocity is a consequence of microtubules with finite length undergoing dynamic instability and is in sharp contrast to the behavior of reaction-diffusion systems, where travelling fronts typically become infinitesimally slow before ceasing to propagate \citep{Hallatschek:2009hm, Mendez:2007ug, vanSaarloos:2003vr, Chang:2013bc}. One can understand the origin of $V_{gap}>0$ when microtubules are eliminated after a catastrophe event ($f_{res}=0; J=-v_{s}$). In this limit, plus ends always expand with the velocity $v_{g}$ until they eventually collapse. Below $r_{c}$, this forward expansion of plus ends fails to produce aster growth because the number of plus ends declines on average. Right above $r_{c}$, the number of plus ends is stable and aster grows at the same velocity as individual microtubules. Indeed, Eq. \eqref{eq:vgap} predicts that $V_{gap}=v_{g}$ when $f_{res}=0$. The dynamics are similar for $f_{res}>0$. At the transition, nucleation stabilizes a subpopulation of microtubules expanding forward, and their average velocity sets the value of $V_{gap}$. We also find that the magnitude of $V_{gap}$ is inversely proportional to the mean length of microtubules in the system (SI). Thus, the shorter the microtubules, the more explosive this transition becomes.

In the SI, we also show that microtubule density inside the aster is proportional to $r-r_{c}$. Thus, the density is close to zero during the transition from stationary to growing asters, but quickly increases as the nucleation rate becomes larger. As a result, cells can achieve rapid aster growth while keeping the density of the resulting microtubule network sufficiently low. The low density might be beneficial because of its mechanical properties or because it simply requires less tubulin to produce and energy to maintain. In addition, the explosive transition to growth with $V_{gap}>0$ allows the cell to independently control the aster density and growth speed.

Model parameters other than the nucleation rate can also be tuned to transition asters from growth to no growth regimes. Similar to Eq. \eqref{eq:rcritical} and \eqref{eq:vgap}, one can define the critical parameter value and gap velocity to encompass all such transitions (Table S1). In all cases, we find that the onset of aster growth is accompanied by discontinuous increase in the growth velocity. The finite jump in aster growth velocity is similarly predicted in a wide range of alternative scenarios including (i) feedback regulation of plus end dynamics (SI and Figure 3 - figure supplement 1) and (ii) aster growth by microtubule polymer-stimulated nucleation (SI and Figure 3 - figure supplement 2). In summary, the gap velocity is a general prediction of the collective behavior of microtubules that are short-lived.

\subsection*{Titration of MCAK Slows then Arrests Aster Growth with Evidence for a Gap Velocity}

Based on our theory, we reasoned that it would be possible to transform a growing interphase aster to a small, stationary aster by tuning polymerization dynamics and/or nucleation via biochemical perturbations in \textit{Xenopus} egg extract. To this end, we performed reconstitution experiments in undiluted interphase cytoplasm supplied with anti-Aurora kinase A antibody coated beads, which nucleate microtubules and initiate aster growth under the coverslip \citep{Ishihara:2014ih, Field:2014eb}. We explored perturbation of various regulators for plus end dynamics and nucleation. We settled on perturbation of MCAK/KIF2C, classically characterized as the main catastrophe-promoting factor in the extract system \citep{Walczak:1996vi, Kinoshita:2001kc}, and imaged aster growth. 

In control reactions, aster radius, visualized by the plus end marker EB1-mApple, increased at velocities of $20.3 \pm 3.1 \mu m/min$ (n=21 asters). We saw no detectable changes to aster growth with addition of the wild type MCAK protein. In contrast, addition of MCAK-Q710 \citep{Moore:2004cf} decreased aster growth velocity (Fig. 4A and B). At concentrations of MCAK-Q710 above 320 nM, most asters had small radii with very few microtubules growing from the Aurora A beads. In our model, this behavior is consistent with any change of parameter(s) that reduces the aster growth velocity (Eq. \eqref{eq:velocity}) and arrests growth.

\begin{figure}[ht]
\includegraphics[width=\textwidth]{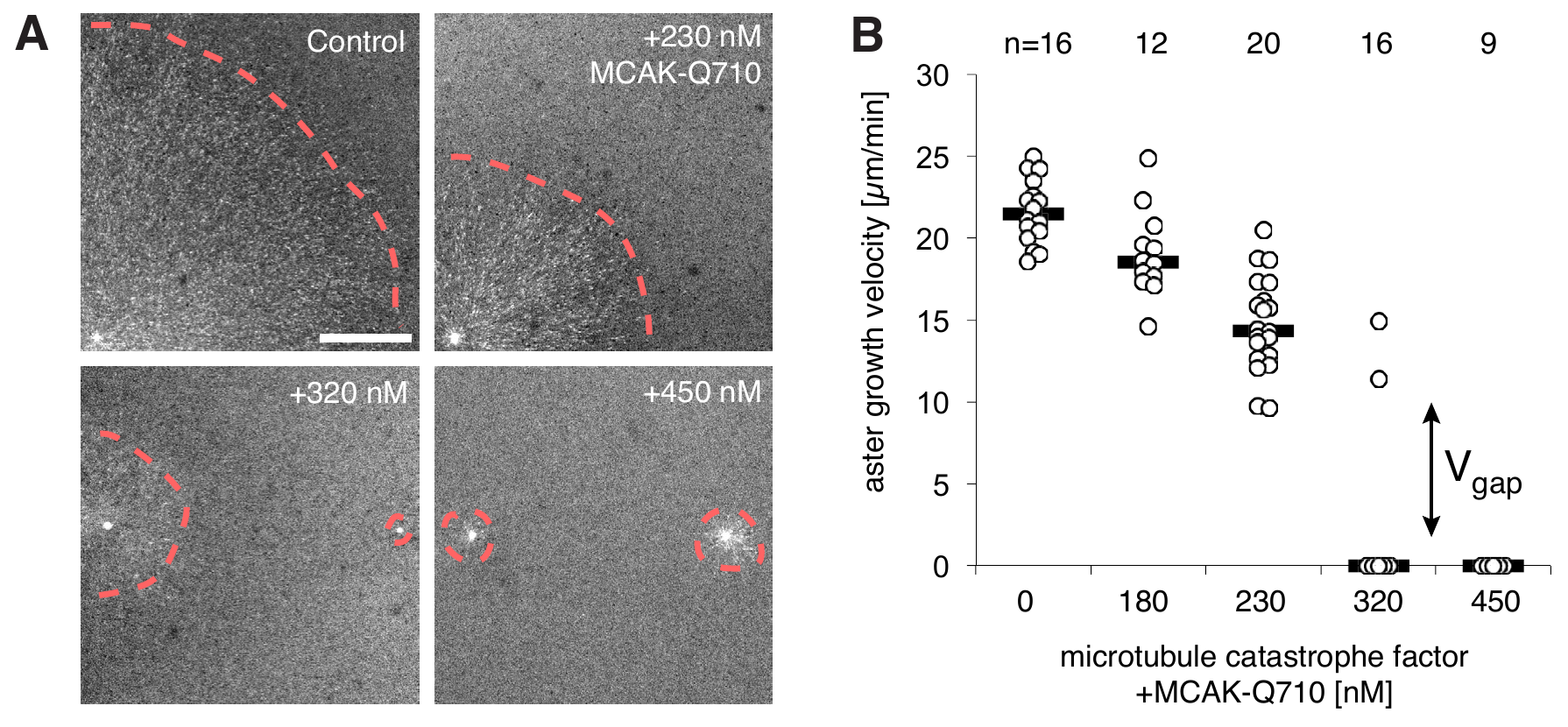}
\caption{\color{Gray} 
\textbf{Titration of MCAK-Q710 slows then arrests aster growth through a discontinuous transition.}
(A) Addition of MCAK-Q710 results in smaller interphase asters assembled by Aurora A beads in \textit{Xenopus} egg extract. Images were obtained 20 minutes post initiation with the plus end marker EB1-mApple. Dotted lines indicate the approximate outline of asters. (B) Aster growth velocity decreases with MCAK-Q710 concentration and then abruptly vanishes as predicted by the model. Note a clear gap in the values of the observed velocities and bimodality near the transition, which support the existence of $V_{gap}$.  Quantification methods are described in methods and Figure 4-figure supplement 1}
\label{MCAK}
\end{figure}

At 320nM MCAK-Q710 concentration, we observed bimodal behavior. Some asters increased in radius at moderate rates, while other asters maintained a stable size before disappearing, presumably due to the decrease of centrosomal nucleation over time (Figure 4-figure supplement 1 and \citep{Ishihara:2014ih}). In particular, we observed no asters growing at velocities between 0 and $9 \mu m/min$ (Fig. 4B and Figure 4-figure supplement 1). This gap in the range of possible velocities is consistent with the theoretical prediction that growing asters expand above a minimal rate $V_{gap}$. 

To confirm that the failure of aster growth at high concentrations of MCAK-Q710 is caused by the changes in aster growth rather than nucleation from the beads, we repeated the experiments with \textit{Tetrahymena} pellicles as the initiating centers instead of Aurora A beads. Pellicles are pre-loaded with a high density of microtubule nucleating sites, and are capable of assembling large interphase asters \citep{Ishihara:2014ih}. We found pellicle initiated asters to exhibit a similar critical concentration of MCAK-Q710 compared to Aurora A bead asters. While the majority of Aurora A beads subjected to the highest concentration of MCAK-Q710 lost growing microtubules over time, a significant number of microtubules persisted on pellicles even after 60 min (Figure 4-figure supplement 2). The radii of these asters did not change, consistent with our prediction of stationary asters. Thus, the pellicle experiments confirmed our main experimental result of small, stationary asters and that the nature of transition is consistent with the existence of a gap velocity.

Finally, we asked which parameters in our model were altered in the MCAK-Q710 perturbation. To this end, we measured the polymerization and catastrophe rates in interphase asters assembled by Aurora A beads at various MCAK-Q710 concentrations. We imaged EB1 comets at high spatiotemporal resolution, and analyzed their trajectories by tracking-based image analysis (\citep{Matov:2010jt, Applegate:2011fn}, Methods). Neither the polymerization nor the catastrophe rate changed at the MCAK-Q710 concentrations corresponding to the transition between growing and stationary asters (Figure 4-figure supplement 3). MCAK-Q710 has been reported to reduce microtubule polymer levels in cells \citep{Moore:2004cf}, but its precise effect on polymerization dynamics and/or nucleation remains unknown. Our data is consistent with the following three scenarios for how MCAK-Q710 antagonizes microtubule assembly: (i) increased depolymerization rate, (ii) decreased rescue rate, and/or (iii) decreased nucleation rate.

\section*{DISCUSSION}

\subsection*{An Autocatalytic Model of Aster Growth}
It has not been clear whether the standard model of aster growth can explain the morphology of asters observed in all animal cells, including those of extreme size \citep{Mitchison:2015bg}. To resolve this question, we constructed a biophysical framework that incorporates microtubule polymerization dynamics and autocatalytic nucleation. Numerical simulations and analytical solutions (Fig. 2, 3, and Fig. 3-figure supplement 1, 2) recapitulated both stationary and continuously growing asters in a parameter-dependent manner. Interestingly, the explosive transition from “growth” to “no growth” was predicted to involve a finite growth velocity, which we confirmed in biochemical experiments (Fig. 4). 

Our biophysical model offers a biologically appealing explanation to aster growth and allows us to estimate parameters that are not directly accessible: the rescue and autocatalytic nucleation rates. For example, if we assume that MCAK-Q710 decreases the nucleation rate, we may use the $V_{gap}$ equation for $r\rightarrow r_c$ (Eq. \eqref{eq:vgap}), the equation for aster growth velocity V (Eq. \eqref{eq:velocity}), and our measurements of $v_g$, $v_s$, $f_{cat}$, $V$, and $V_{gap}$ (Table 1) to simultaneously estimate $f_{res}$ and $r$. These results are summarized in Table 1. Our inferred value of autocatalytic nucleation $r=2.1 \textrm{min}^{-1}$ is comparable to previous estimates: 1.5 $\textrm{min}^{-1}$ \citep{Clausen:2007fi} and 1 $\textrm{min}^{-1}$ \citep{Petry:2013iu} in meiotic egg extract supplemented with RanGTP. In the alternative scenarios, where MCAK-Q710 decreases the catastrophe rate or increases the depolymerization rate, our estimates of $r$ and $f_{res}$ are essentially the same (Table S2). Thus, our model recapitulates aster growth with reasonable parameter values and offers a new understanding for how asters grow to span large cytoplasms even when individual microtubules are unstable.

To date, few studies have rigorously compared the mechanistic consequences of plus-end-stimulated vs. polymer-stimulated nucleation. Above, we presented the theoretical predictions for aster growth by plus-end stimulated nucleation. In the SI, we also provide the results for polymer-stimulated nucleation including the critical nucleation rate Eq. S59. Both models of nucleation have qualitatively similar behavior including the gap velocity and recapitulate experimental observations of asters growing as traveling waves. Thus, in our case, the qualitative conclusions do not depend on the precise molecular mechanism of autocatalytic nucleation. In particular, the explosive transition characterized by the gap velocity is a general prediction of modeling microtubules as self-amplifying elements whose lifetime depends on their length. 

By carefully defining and quantifying autocatalytic nucleation, future studies may be able to distinguish its precise mechanism. With plus-end-stimulated nucleation, the nucleation rate $r$ has units of $\textrm{min}^{-1}$ and describes the number of new microtubules generated per existing plus end per minute. With polymer-stimulated nucleation, the nucleation rate has units of $\mu\textrm{m}^{-1} \textrm{min}^{-1}$, and describes the number of new microtubules generated per micron of existing microtubule per minute. This difference has important implications for the structural mechanism of microtubule nucleation and for the prediction of cell-scale phenomena. For the issue of large aster growth, we propose specific experiments that might be able distinguish these scenarios (SI).

\subsection*{Phase Diagram for Aster Growth}
How do large cells control aster size during rapid divisions? We summarize our theoretical findings with a phase diagram for aster growth in Fig. 5. Small mitotic asters are represented by stationary asters found in the regime of bounded polymerization dynamics $J<0$ and low nucleation rates. These model parameters must change as cells transition from mitosis to interphase to produce large growing asters. Polymerization dynamics becomes more favorable to elongation during interphase \citep{Belmont:1990vs, Verde:1992tl}. This may be accompanied by an increased autocatalytic nucleation of microtubules. 

\begin{figure}[ht]
\includegraphics[width=0.8\textwidth]{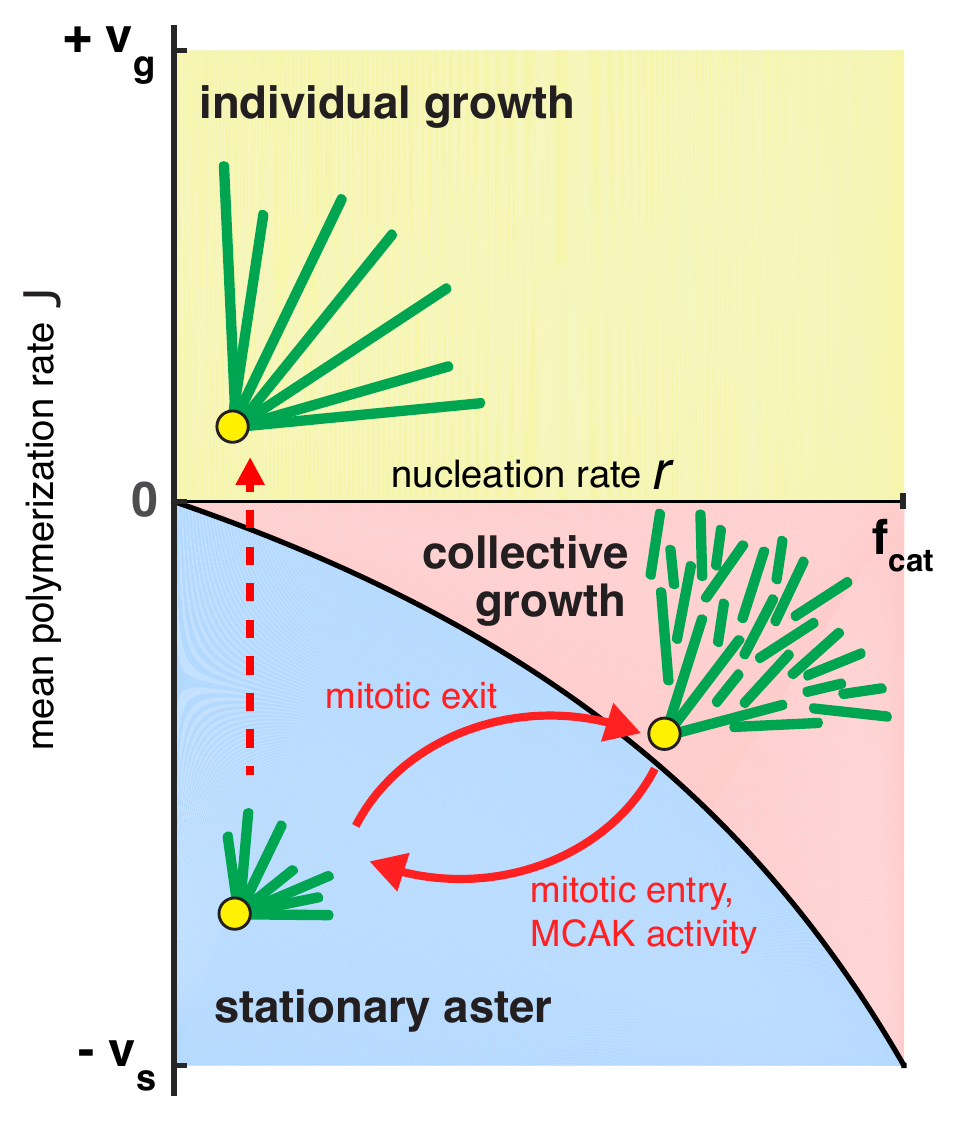}
\caption{\color{Gray} 
\textbf{Phase diagram for aster growth.}
Aster morphology is determined by the balance of polymerization dynamics and autocatalytic nucleation. Small, stationary asters ($V=0$), as observed during mitosis, occur at low nucleation $r$ and net depolymerization of individual microtubules ($J<0$). Net polymerization ($J>0$) without nucleation ($r=0$) produces asters that expand at rate $V=J$ with dilution of microtubule density at the periphery and are thus inconsistent with experimental observations. The addition of nucleation to the individual growth regime changes these dynamics only marginally (yellow region); see SI. Alternatively, the transition from stationary to growing asters can be achieved by increasing the nucleation rate, $r$, while keeping $J$ negative. Above the critical nucleation rate $r_{c}$ starts the regime of collective growth ($V$ as in Eq. \eqref{eq:velocity}, which is valid for $r_{c}<r<f_{cat}$) that produces asters composed of relatively short microtubules (red region). The transition from stationary aster to collective growth may be achieved by crossing the curve at any location, but always involves an explosive jump in aster growth velocity, $V_{gap}$. Reverse transition recapitulates the results of our experimental perturbation of MCAK activity (Fig. 4) and mitotic entry (solid arrows). We propose this unified biophysical picture as an explanation for the cell cycle dependent changes of aster morphology \textit{in vivo}.
}
\label{phase}
\end{figure}

According to the standard model, increasing $J$ to a positive value with no nucleation leads to asters in the “individual growth” regime. A previous study suggested the interphase cytoplasm is in the unbounded polymerization dynamics $J>0$ \citep{Verde:1992tl}, but our measurements of parameters used to calculate $J$ differ greatly (Table 1). Individual growth regime is also inconsistent with the steady-state density of microtubules at the periphery of large asters in both fish and frog embryos \citep{Wuhr:2008dz, Wuhr:2010fi, Ishihara:2014ih}. Experiments in egg extracts further confirm the addition of new microtubules during aster growth \citep{Ishihara:2014ih} contrary to the predictions of the standard model. Furthermore, the presence of a high density of growing plus ends in the interior of growing asters in egg extract suggests that microtubules must be short compared to aster radius, and mean growth velocity must be negative, at least in the aster interior \citep{Ishihara:2014ih}. 

By constructing a model that incorporates autocatalytic nucleation $r>0$, we discovered a new regime, in which continuous aster growth is supported even when microtubules are unstable ($J<0$). We call this the “collective growth” regime because individual microtubules are much shorter (estimated mean length of 20 $\mu$m, Table 1) than the aster radius (hundreds of microns). Predictions of this model are fully confirmed by the biochemical perturbation via MCAK-Q710. The finite jump in the aster growth velocity (Fig. 4) is in sharp contrast to the prediction of the standard models of spatial growth \citep{Fisher:1937hy, Kolmogorov:1937tw, Skellam:1951em, vanSaarloos:2003vr}. Spatial growth is typically modeled by reaction-diffusion processes that account for birth events and random motion, which, in the context of microtubules, correspond to the nucleation and dynamic instability of plus ends. Reaction-diffusion models, however, neglect internal dynamics of the agents such as the length of a microtubule. As a result, such models inevitably predict a continuous, gradual increase in the growth velocity as the model parameters are varied \citep{Hallatschek:2009hm, Mendez:2007ug, vanSaarloos:2003vr, Chang:2013bc}. The observation of finite velocity jump provides a strong support for our model and rules out a very wide class of models that reproduce the overall phenomenology of aster growth including the constant velocity and profile shape (Fig. 2). In particular, the observation of $V_{gap}$ excludes the model that we previously proposed based on the analogy of aster growth and the Fisher-Kolmogorov equation \citep{Ishihara:2014gpa}. The implications of $V_{gap}$ for model selection are further discussed in SI.

\subsection*{Collective Growth of Cytoskeletal Structures}
Our theory allows for independent regulation of aster growth rate and microtubule density through the control of the nucleation rate and microtubule polymerization. Thus, cells have a lot of flexibility in optimizing aster properties and behavior. The existence of a gap velocity results in switch-like transition from quiescence to rapid growth and allows cells to drastically alter aster morphology with a small change of parameters. Importantly, the rapid growth does not require high microtubule density inside asters, which can be tuned independently. 

Collective growth produces a meshwork of short microtubules with potentially desirable properties. First, the network is robust to microtubule severing or the spontaneous detachment from the centrosome.  Second, the network can span arbitrarily large distances yet disassemble rapidly upon mitotic entry. Third, the structure, and therefore the mechanical properties, of the network do not depend on the distance from the centrosome. As a speculation, the physical interconnection of the microtubules may facilitate the transduction of mechanical forces across the cell in a way unattainable in the radial array predicted by the standard model \citep{Tanimoto:2016db, Wuhr:2010fi}. 

The regime of collective growth parallels the assembly of other large cellular structures from short, interacting filaments \citep{Pollard:2003vg} and is particularly reminiscent of how meiosis-II spindles self-assemble \citep{Burbank:2007jx, Brugues:2012fx, Brugues:2014bp}. Due to such dynamic architecture, spindles are known to have unique physical properties such as self-repair, fusion \citep{Gatlin:2009hq} and scaling \citep{Wuhr:2008dz, Hazel:2013ig, Good:2013dc}, which could allow for greater robustness and evolvability \citep{Kirschner:1998tp}. Perhaps, collective growth is one of the most reliable ways for a cell to assemble cytoskeletal structures that exceed the typical length scales of individual filaments. 

\section*{MATERIALS AND METHODS}

\subsection*{Numerical Simulations}
We implemented a finite difference method with fixed time steps to numerically solve the continuum model (Eq. \eqref{eq:PDEmodel}). Forward Euler’s discretization scheme was used except exact solutions of advection equations was used to account for the gradient terms. Specifically, the plus end positions were simply shifted by $+v_{g}\delta t$ for growing microtubules and by $-v_{s}\delta t$ for shrinking microtubules. Nucleation added new growing microtubules of zero length at a position-dependent rate given by $Q(x)$. The algorithm was implemented using MATLAB (Mathworks).

\subsection*{Analytical Solution}
We linearized Eq. \eqref{eq:PDEmodel} for small $C_{g}$ and solved it using Laplace transforms in both space and time. The inverse Laplace transform was evaluated using the saddle point method \citep{Bender:1999uu}. We found the aster growth velocity as in Eq. \eqref{eq:velocity}. The details of this calculation are summarized in the Supporting Text (SI).

\subsection*{Aster Growth Velocity Measurements}
Interphase microtubule asters were reconstituted in \textit{Xenopus} egg extract as described previously with use of p150-CC1 to inhibit dynein mediated microtubule sliding \citep{Field:2014eb, Ishihara:2014ih}. Fluorescence microscopy was performed on a Nikon 90i upright microscope equipped with a Prior Proscan II motorized stage. EB1-mApple was imaged every 2 min with a 10x Plan Apo 0.45 N.A. or a 20x Plan Apo 0.75 N.A. objective. For the analysis of the aster growth front, a linear region originating from the center of asters was chosen (Figure 4-figure supplement 1). A low pass filter was applied to the fluorescence intensity profile and the half-max position, corresponding to the aster edge, was determined manually. The analysis was assisted by scripts written in ImageJ and MATLAB (Mathworks). Univariate scatter plots were generated with a template from \citep{Weissgerber:2015er}. EB1-mApple were purified as in \citep{Petry:2011bi}, used at a final concentration of 100 nM. In some experiments, MCAK or MCAK-Q710-GFP \citep{Moore:2004cf} proteins were added to the reactions. Protein A Dynabeads coated with anti-Aurora kinase A antibody \citep{Tsai:2005ja} or \textit{Tetrahymena} pellicles were used as microtubule nucleating sites.

\subsection*{Catastrophe Rate Measurements}
Interphase asters were assembled as described above. Catastrophe rates and plus end polymerization rates were estimated from time lapse images of EB1 comets that localize to growing plus ends \citep{Matov:2010jt}.  The distributions of EB1 track durations were fitted to an exponential function to estimate the catastrophe rate. Spinning disc confocal microscopy was performed on a Nikon Ti motorized inverted microscope equipped with Perfect Focus, a Prior Proscan II motorized stage, Yokagawa CSU-X1 spinning disk confocal with Spectral Applied Research Aurora Borealis modification, Spectral Applied Research LMM-5 laser merge module with AOTF controlled solid state lasers: 488nm (100mW), 561nm (100mW), and Hamamatsu ORCA-AG cooled CCD camera. EB1-mApple was imaged every 2 sec with a 60x Plan Apo 1.40 N.A. objective with 2x2 binning. EB1 tracks were analyzed with PlusTipTracker \citep{Applegate:2011fn}.

\section*{Author Contributions}
KI and KK developed and analyzed the model. KI performed the experiments and analyzed the data. KI, KK, and TM designed the research and wrote the manuscript.

\section*{Acknowledgments}
We thank the members of Mitchison and Korolev groups for helpful discussion. This work was supported by NIH grant GM39565 and by MBL summer fellowships. The computations in this paper were run on the Odyssey cluster supported by the FAS Division of Science, Research Computing Group at Harvard University. We thank Nikon Imaging Center at Harvard Medical School, and Nikon Inc. at Marine Biological Laboratory for microscopy support. We thank Ryoma Ohi for providing MCAK and expert advice. We thank Linda Wordeman for providing MCAK-Q710-GFP and expert advice. KK was supported by a start up fund from Boston University and by a grant from the Simons Foundation $\#$409704. KI was supported by the Honjo International Scholarship Foundation.

\clearpage

\begin{table}[!ht]
\begin{adjustwidth}{-2.25in}{0in}
\begin{tabular}{lccl}
Quantity & Symbol & Value & Comment\tabularnewline
\hline 
Polymerization rate & $v_{g}$ & 30 $\mu$m/min & Measured from growing plus ends and EB1 comets \tabularnewline
Depolymerization rate & $v_{s}$ & 42 $\mu$m/min & Measured from shrinking plus ends \citep{Ishihara:2014ih} \tabularnewline
Catastrophe rate & $f_{cat}$ & 3.3 min$^{-1}$ & Measured from EB1 comet lifetimes (see Methods)\tabularnewline
Rescue rate & $f_{res}$ & 2.0$\pm$0.3 min$^{-1}$ & Estimated from Eq. \eqref{eq:velocity} and \eqref{eq:vgap} \tabularnewline
Autocatalytic nucleation rate & $r$ & 2.1$\pm$0.2 min$^{-1}$ & Estimated from Eq. \eqref{eq:velocity} and \eqref{eq:vgap} \tabularnewline
Carrying capacity of growing ends & $K$ & 0.4 $\mu\textrm{m}^{-2}$ & Estimated from comparing $C_{g}^{bulk}$ to predicted (see SI)\tabularnewline
Mean microtubule length & $\langle l\rangle$ & 16 $\pm$ 2 $\mu\textrm{m}$ & Estimated from from dynamics parameters (see SI) \tabularnewline
Aster growth velocity & $V$ & 22.3$\pm$2.6 $\mu$m/min & Measured from rate of aster radius increase \tabularnewline
Gap velocity & $V_{gap}$ & 12.8$\pm$1.7 $\mu$m/min & Measured from aster growth at 320 nM MCAK-Q710 \tabularnewline
Bulk growing plus end density & $C_{g}^{bulk}$ & 0.053$\pm$0.030 $\mu\textrm{m}^{-2}$ & Measured from EB1 comet density \citep{Ishihara:2014ih} \tabularnewline
\hline 
\end{tabular}
\protect\caption{Model parameters used to describe large aster growth reconstituted in interphase \textit{Xenopus} egg extract.}
\end{adjustwidth}
\end{table}

\clearpage
\setcounter{equation}{0}


\renewcommand{\theequation}{S\arabic{equation}}

\makeatother


\part*{Appendix: A continuum description of aster growth}

\tableofcontents{}

\section{System of microtubules undergoing plus end polymerization dynamics
and autocatalytic nucleation}

To describe our system of microtubules, we define the mean-field variable
$\rho_{g}(t,x,l)$ which represents the local density of growing plus
ends of length $l$ with its corresponding minus end at position $x$
at time $t$. Similarly, we define the shrinking plus end density
$\rho_{s}(t,x,l)$. We assume that all microtubules have the same
polarity, namely that all microtubules have their plus ends pointing
outwards. Given this polarity, the plus end position of a microtubule
is $x+l$. We define nucleation as the birth of length zero growing
microtubules. The nucleation rate is denoted by $Q_{(t,x)}$. The
evolution of our system is described as follows:

\begin{equation}
\left\{ \ \begin{aligned}\frac{\partial\rho_{g}}{\partial t}=-v_{g}\frac{\partial\rho_{g}}{\partial l}-f_{cat}\rho_{g}+f_{res}\rho_{s} & +Q\cdot\delta(l),\\
\frac{\partial\rho_{s}}{\partial t}=+v_{s}\frac{\partial\rho_{s}}{\partial l}+f_{cat}\rho_{g}-f_{res}\rho_{s,}
\end{aligned}
\right.\label{eq:PDEsystem}
\end{equation}

where $\delta(l)$ is the Dirac delta function.%
\footnote{Instead of a delta function as in Eq. (\ref{eq:PDEsystem}), one can
introduce nucleation as a boundary condition specifying the flux of
new microtubules into the system, $Q(t,x)=v_{g}\rho_{g}(t,\, x,\, l=0)$.%
} The above expression represents an infinite set of equations valid
for the continuum $x\geq0$.

In general, the nucleation rate may depend on $\rho_{g}$ and $\rho_{s}$.
In the following sections, we assume that nucleation proceeds as bifurcation
of growing plus ends.

Nucleation is expressed as a logistic function of local growing plus
end density with carrying capacity of the system as $K$. %
\footnote{An alternative mechanism for autocatalytic nucleation is a scenario
where the local polymer density stimulates nucleation. This may better
relate to proposed models in which freely diffusing nucleation complexes
bind to the side of pre-existing microtubules and become activated.
The nucleation term in Eq. (\ref{eq:nucleationbifurcation}) has the
advantage that the exact asymptotic solution may be solved and we
argue that it captures the qualitative effect of autocatalytic nucleation.%
}

\begin{equation}
Q(t,x)=v_{g}\,\rho_{g}(t,\, x,\, l=0)=r\cdot C_{g}(t,x)\left(1-\dfrac{{C_{g}(t,x)}}{K}\right),\label{eq:nucleationbifurcation}
\end{equation}

where $C_{g}(t,x_{+})$ denotes the local growing plus end density
at position $x_{+}$ and time $t$. By equating the plus end density
times the surface area of a sphere of radius $x_{+}$ to the integration
of $\rho_{g}(t,x,l)$ whose plus ends happens be positioned at distance
$x_{+}$ , we obtain the following expression for local growing plus
end density:

\begin{equation}
C_{g}(t,x)=\int_{0}^{+\infty}dl\int_{0}^{x}dx_{-}\,\rho_{g}(t,x_{-},l)\cdot\delta(x-x_{-}-l)\label{eq:deflocalgrowingends}
\end{equation}

Similarly, we define the shrinking plus end density $C_{s}(t,x_{+})$.

In higher dimensions, Eq. (\ref{eq:deflocalgrowingends}) generalizes
as follows

\begin{equation}
C_{g}(t,|x|)=\dfrac{1}{|x|^{d-1}}\int_{0}^{+\infty}dl\int_{0}^{|x|}d|x_{-}|\,\rho_{g}(t,|x_{-}|,l)\cdot\delta(|x|-|x_{-}-l|)\,|x|^{d-1}\textrm{,}\label{eq:higherdimensions}
\end{equation}

where we have assumed spherical geometry.

\section{Bounded and unbounded regimes of polymerization dynamics}

In the absence of nucleation ($r=0$), the dynamics of our system
is goverened solely by microtubule plus end dynamics. One distinguishes
two tupes of growth: bounded and unbounded. In the former case, the
mean polymerization rate $J=\dfrac{v_{g}f_{res}-v_{s}f_{cat}}{f_{res}+f_{cat}}$
is negative and microtubules shrink on average. In the latter case,
$J$ is positive and microtubule become progressively longer. $J$
is essentially the directional bias of the plus end dynamics, the
drift term of the biased random walk \cite{Dogterom:1993tz,Verde:1992tl,Bicout:1997tz}.

Imagine a scenario where there is a fixed number of microtubules in
the system, and that all their minus ends are at the origin. Further,
let us assume that when plus ends shrink back to their minus ends,
they instantly transition to a growing state, akin to a reflective
boundary condition at the origin. When $J<0$, the system will reach
a steady-state where the length of individual microtubules are found
to be exponentially distributed with an average length of $\langle l\rangle=\dfrac{-v_{g}v_{s}}{v_{g}f_{res}-v_{s}f_{cat}}$.
When $J>0,$ eventually all microtubules will be long enough with
their plus ends far from the origin. Thus, there is no steady-state
length distribution and the average length increases at rate $J$.

\section{Aster growth dynamics with autocatalytic microtubule nucleation}

Here, we analyze the behavior of the system when polymerization dynamics
and autocatalytic microtubule nucleation are allowed. Our intuition
is that microtubule nucleation will produce microtubules at distances
far from the origin, and that with high enough nucleation, the population
of microtubules will start to move away from the origin as a self-propagating
wave. Waves with growth terms that monotonically decline with density
are called pulled fronts, and can be analyzed through linearization
\cite{vanSaarloos:2003vr}. This is the case for the microtubule nucleation
specified by Eq. (\ref{eq:nucleationbifurcation}), and the full analytical
solution for the spatiotemporal evolution can be obtained. In particular,
we find the asymptotic front velocity of microtubules as a function
of model parameters. This leads us to the concept of the critical
nucleation rate, which summarizes the condition that determines aster
growth (linear increase in radius) vs. no growth (constant radius).
Other feedback mechanisms in $r,\, f_{cat},v\,_{g},\textrm{\,\ etc.}$
lead to the same final equations where all parameters are specified
at low density (see section \ref{sub:Other-types-of} for an extended
discussion).

\subsection{Solution of the system}

We apply the Laplace transform to Eq. (\ref{eq:PDEsystem}) in the
following way: time domain $t\rightarrow s$, spatial domain $x_{-}\rightarrow k$
and length domain $l\rightarrow q$. Our system recast in $\rho_{g}(s,k,q)$
and $\rho_{s}(s,k,q)$ becomes

\begin{equation}
\left\{ \ \begin{aligned}s\rho_{g}-\rho_{g}(t=0,k,q)=-v_{g}q\rho_{g}+v_{g}\rho_{g}(s,k,l=0)-f_{cat}\rho_{g}+f_{res}\rho_{s},\quad\quad\quad\\
s\rho_{s}-\rho_{s}(t=0,k,q)=+v_{s}q\rho_{s}-v_{s}\rho_{s}(s,k,l=0)+f_{cat}\rho_{g}-f_{res}\rho_{s}.\quad\quad\quad
\end{aligned}
\right.\label{eq:LaplacedPDE}
\end{equation}

We can not directly solve the system Eq. (\ref{eq:LaplacedPDE}) for
$\rho_{g}(s,k,q)$ and $\rho_{s}(s,k,q)$ as $\rho_{g}(s,k,l=0)$
and $\rho_{s}(s,k,l=0)$ are unknown. However, we demonstrate that
the system can be closed for the local plus end densities $C_{g}(s,k)$
and $C_{s}(s,k)$ when the nucleation rate depends on $\rho_{g}(t,k,l)$
and $\rho_{s}(t,k,l)$ as in Eq. (\ref{eq:nucleationbifurcation}).
We substitute this solution back to Eq. (\ref{eq:LaplacedPDE}) and
obtain the full solution of the system in terms of $\rho_{g}$ and
$\rho_{s}$.

First, let us consider the transformation of $C_{g}(t,x)$ in the
spatial domain $x\rightarrow k$.
\begin{align}
C_{g}(t,k) & =\int_{0}^{+\infty}e^{-kx}dx\, C_{g}(t,x)\nonumber \\
 & =\int_{0}^{+\infty}\, e^{-kx_{+}}dx\,\int_{0}^{+\infty}dl\int_{0}^{+\infty}dx_{-}\,\delta(x-x_{-}-l)\,\rho_{g}(t,x,l)\label{eq:CRhotransform}\\
 & =\int_{0}^{+\infty}\int_{0}^{+\infty}dx\, dl\ \rho_{g}(t,x,l)\ e^{-kx}e^{-kl}\nonumber \\
 & =\int_{0}^{+\infty}dl\,\rho_{g}(t,k,l)\ e^{-kl}\nonumber \\
 & =\rho_{g}(t,k,q=k)\nonumber 
\end{align}

We apply the same transform to $C_{s}(t,x_{+})$ and obtain the following:

\begin{equation}
C_{g}(s,k)=\rho_{g}(s,k,k)\quad\textrm{and}\quad C_{s}(s,k)=\rho_{s}(s,k,k)\label{eq:Crhoequivalence}
\end{equation}

Thus, we find that the local plus end density $C_{g}(s,k)$ is equivalent
to a special subset of $\rho_{g}(s,k,q)$ that is $\rho_{g}(s,k,k)$,
where the spatial and length domains are coupled. By applying Eq.
(\ref{eq:Crhoequivalence}) and $k=q$ to Eq. (\ref{eq:LaplacedPDE}),
we obtain

\begin{equation}
\left\{ \ \begin{aligned}sC_{g}-C_{g}(t=0,k)=-v_{g}kC_{g}+v_{g}\rho_{g}(l=0)-f_{cat}C_{g}+f_{res}C_{s},\\
sC_{s}-C_{s}(t=0,k)=+v_{s}kC_{s}-v_{s}\rho_{s}(l=0)+f_{cat}C_{g}-f_{res}C_{s}.
\end{aligned}
\right.\label{eq:transformedPDE}
\end{equation}

Note that the term corresponding to growing plus ends at $l=0$ is
equivalent to our defintion of nucleation rate Eq. (\ref{eq:nucleationbifurcation}).
Here, we substitute the linearized form of the boundary condition
with Eq. (\ref{eq:nucleationbifurcation}), $v_{g}\,\rho_{g}(s,\, k,\, l=0)=r\cdot C_{g}(s,k)$,
which is valid for small $C_{g}$ or at the leading edge of the aster.
The term corresponding to shrinking plus ends at $l=0$ is not directly
specified in our system. The value of $v_{s}\rho_{s}(l=0)$ can be
obtained from a special condition required for physical consistency
as we show below. For now, we will treat $\rho_{s}(l=0)$ as known.

From the initial conditions $\rho_{g}^{0}=\rho_{g}(t=0,k,q)$ and
$\rho_{s}^{0}=\rho_{s}(t=0,k,q)$, we know $C_{g}(t=0,k)=\rho_{g}(t=0,k,k)$
and $C_{s}(t=0,k)=\rho_{s}(t=0,k,k)$. We arrive at the subproblem:

\begin{equation}
\begin{array}{c}
A_{r}\left(\begin{array}{c}
C_{g}(s,k)\\
C_{s}(s,k)
\end{array}\right)=\left(\begin{array}{c}
\rho_{g}(t=0,k,k)\\
\rho_{s}(t=0,k,k)-v_{s}\rho_{s}(l=0)
\end{array}\right)\textrm{,}\\
\textrm{where}\ A_{r}=\left[\begin{array}{cc}
s+v_{g}k+f_{cat}-r & -f_{res}\\
-f_{cat} & s-v_{s}k+f_{res}
\end{array}\right]\textrm{.}
\end{array}\label{eq:MatrixSystem1}
\end{equation}

This system is solved by matrix inversion. The solution for growing
plus end density is

\begin{equation}
C_{g}(s,k)=\dfrac{1}{\textrm{det}(A_{r})}\left[A_{r,22}\rho_{g}(t=0,k,k)-A_{r,12}(\rho_{s}(t=0,k,k)-v_{s}\rho_{s}(l=0))\right]\textrm{.}\label{eq:SolutionCg}
\end{equation}

With the knowledge of $C_{g}(s,k)$, we proceed to find the general
solution for $\rho_{g}(s,k,q)$ by returning to the full problem Eq.
(\ref{eq:PDEsystem}). Arranging the known quantitites to the right
hand side, the rewritten problem reads

\begin{equation}
\begin{array}{c}
A\left(\begin{array}{c}
\rho_{g}(s,k,q)\\
\rho_{s}(s,k,q)
\end{array}\right)=\left(\begin{array}{c}
\rho_{g}^{0}+rC_{g}(s,k)\\
\rho_{s}^{0}-v_{s}\rho_{s}(l=0)
\end{array}\right)\textrm{,}\\
\textrm{where}\ A=\left[\begin{array}{cc}
s+v_{g}q+f_{cat} & -f_{res}\\
-f_{cat} & s-v_{s}q+f_{res}
\end{array}\right]\textrm{.}
\end{array}\label{eq:MatrixSystem2}
\end{equation}

For the solution of plus end density, we substitute our solution of
$C_{g}(s,k)$ and obtain

\begin{equation}
\begin{aligned}\rho_{g}(s,k,q) & =\dfrac{1}{\mathrm{det}(A)}\left[A_{22}(\rho_{g}^{0}+rC_{g}(s,k))-A_{12}(\rho_{s}^{0}-v_{s}\rho_{s}(l=0))\right]\\
 & \begin{aligned}\begin{aligned}= & \dfrac{1}{\mathrm{det}(A)}\left(A_{22}\rho_{g}^{0}-A_{12}\rho_{s}^{0}\right)+\dfrac{A_{12}}{\mathrm{det}(A)}v_{s}\rho_{s}(l=0)+\\
 & \dfrac{1}{\textrm{det}(A)}\dfrac{rA_{22}}{\textrm{det}(A_{r})}\left(A_{r,22}\rho_{g}(t=0,k,k)-A_{r,12}\rho_{s}(t=0,k,k)+A_{r,12}v_{s}\rho_{s}(l=0)\right)
\end{aligned}
\end{aligned}
\\
 & \begin{aligned}= & \dfrac{1}{\mathrm{det}(A)}\left(A_{22}\rho_{g}^{0}-A_{12}\rho_{s}^{0}\right)+v_{s}\rho_{s}(l=0)(\dfrac{A_{12}}{\mathrm{det}(A)}+\dfrac{rA_{22}A_{r,12}}{\mathrm{det}(A)\mathrm{det}(A_{r})})\\
 & +\dfrac{rA_{22}}{\mathrm{det}(A)\textrm{det}(A_{r})}(A_{r,22}\rho_{g}(t=0,k,k)-A_{r,12}\rho_{s}(t=0,k,k))\textrm{.}
\end{aligned}
\end{aligned}
\label{eq:solutionRhoG}
\end{equation}

\subsection{Physical plausibility at $l\rightarrow\infty$}

In Eq. (\ref{eq:solutionRhoG}), we still have one unknown $\rho_{s}(l=0)$,
which we determine by imposing a physical plausibility condition.
Namely, we impose that the number of infinitely long microtubules
are zero by requesting that the solution $\rho_{g}$ (and $\rho_{s}$)
decays as $l\rightarrow\infty$. To obtain $\rho_{s}(l=0)$, we apply
the inverse Laplace transform $q\rightarrow l$,

\begin{equation}
\begin{aligned}\rho_{g}(s,k,l) & =\dfrac{1}{2\pi i}\int_{\gamma-i\infty}^{\gamma+i\infty}e^{ql}dq\ \rho_{g}(s,k,q)\\
 & =\sum\underset{q_{j}}{\textrm{Res}}\ e^{q_{j}l}\rho_{g}(s,k,q_{j})
\end{aligned}
\label{eq:InvLaplacesolutionRhoG}
\end{equation}

Note that the only term that depends on $q$ and can give rise to
a residue in Eq. (\ref{eq:solutionRhoG}) is $\textrm{det}(A)$. The
condition $\mathrm{det}(A)=0$ is equivalent to 

\begin{equation}
q^{2}v_{g}v_{s}-q[s(v_{g}-v_{s})+v_{g}f_{res}-v_{s}f_{cat}]-[s^{2}+s(f_{cat}+f_{res})]=0\label{eq:quadraticForQ}
\end{equation}

This equation has two roots $q_{+}>0$ and $q_{-}<0$, since

\begin{equation}
q_{+}q_{-}=-\dfrac{s^{2}+s(f_{cat}+f_{res})}{v_{g}v_{s}}<0\textrm{.}\label{eq:product_qplusqpminus}
\end{equation}
Therefore, the $\rho_{g}(s,k,l)$ takes the form $\rho_{g}(s,k,l)=e^{q_{+}l}\rho_{g}(s,k,q_{+})+e^{q_{-}l}\rho_{g}(s,k,q_{-})$.

We require that $\rho_{g}(s,k,l)$ takes a finite value in the limit
of $l\rightarrow\infty$ by setting the the coefficient for $e^{q_{+}l}$
to zero, in other words $\rho_{g}(s,k,q_{+})=0$. Thus, with $A^{+}$
denoting matrix $A$ when $q=q_{+}$ and $\rho_{g}^{0,+}=\rho_{g}(t=0,k,q^{+})$,
we request the following condition:

\begin{equation}
\begin{aligned}A_{22}^{+}\rho_{g}^{0,+}-A_{12}^{+}\rho_{s}^{0,+}+v_{s}\rho_{s}(l=0)(A_{12}^{+}+\dfrac{rA_{22}^{+}A_{r,12}}{\mathrm{det}(A_{r})})\\
+\dfrac{rA_{22}^{+}}{\textrm{det}(A_{r})}(A_{r,22}\rho_{g}(t=0,k,k)-A_{r,12}\rho_{s}(t=0,k,k))=0
\end{aligned}
\label{eq:RhoS_L0condition}
\end{equation}

Recall that $\rho_{s}(l=0)$ was introduced as an unknown in our system
via Laplace transform despite the absence of a $bona\, fide$ boundary
condition for shrinking microtubules of length zero. We may now solve
the above equation%
\footnote{Similar arguments were previously used by \cite{Bicout:1997tz} in
the context of bounded, non-expanding, asters. %
} for $\rho_{s}(l=0)$, and substitute it to Eq. (\ref{eq:solutionRhoG}),
resulting in our final solution

\begin{equation}
\begin{aligned}\rho_{g}(s,k,q)= & \frac{A_{22}\rho_{g}^{0}-A_{12}\rho_{s}^{0}}{\textrm{det}(A)}+\dfrac{rA_{22}(A_{r,22}\rho_{g}(t=0,k,k)-A_{r,12}\rho_{s}(t=0,k,k))}{\textrm{det}(A)\textrm{det}(A_{r})}\\
 & +\frac{1}{\textrm{det}(A)\textrm{det}(A_{r})}\cdot\dfrac{A_{12}\textrm{det}(A_{r})+rA_{22}A_{r,12}}{A_{12}^{+}\textrm{det}(A_{r})+rA_{22}^{+}A_{r,12}}\cdot\Big(-\rho_{g}^{0,+}A_{22}^{+}\textrm{det}(A_{r})\\
 & \,-\rho_{g}(t=0,k,k)rA_{22}^{+}A_{r,22}+\rho_{s}^{0,+}A_{12}^{+}\textrm{det}(A_{r})+\rho_{s}(t=0,k,k)rA_{22}^{+}A_{r,12}\Big)\textrm{.}
\end{aligned}
\label{eq:solutionRhoGfinal}
\end{equation}

\subsection{Summary of solutions $C_{g}(s,k)$ and $\rho_{g}(s,k,q)$}

Assuming of microtubule nucleation as in equations (\ref{eq:nucleationbifurcation})
and (\ref{eq:deflocalgrowingends}), we have derived the full solution
for $\rho_{g}(s,k,q)$ written as Eq. (\ref{eq:solutionRhoGfinal}).
The solution for $\rho_{s}(s,k,q)$ may be derived similarly. These
solutions directly correspond to the plus end density solutions $C_{g}(s,k)=\rho_{g}(s,k,k)$
and $C_{s}(s,k)=\rho_{s}(s,k,k)$. Direct experimental measurements
are available for $C_{g}(t,x)$. As we are primarly interested in
the velocity at which the front of plus ends advance in the long time
limit, we proceed with our analysis focusing on the behavior of $C_{g}(t,x)$.

\section{Critical transition to aster growth}

\subsection{Dispersion relations for $C_{g}(t,x)$}

We wish to determine the velocity at which the aster expands it radius.
We apply the inverse Laplace transform $s\rightarrow t$ to $C_{g}(t,k)$

\begin{equation}
\begin{aligned}\begin{aligned}C_{g}(t,k)\end{aligned}
 & =\dfrac{1}{2\pi i}\int_{\gamma-i\infty}^{\gamma+i\infty}e^{st}ds\ C_{g}(s,k)\\
 & =\sum_{j}\textrm{Res}\, C_{g}(s_{j}(k),k)\, e^{s_{j}(k)t}
\end{aligned}
\textrm{,}\label{eq:inverseLaplace}
\end{equation}

\noindent where the subscript $j$ specifies the different poles in
$C_{g}(s,k)$. Further, we apply the inverse Laplace transform $k\rightarrow x,$

\begin{equation}
\ \begin{aligned}C_{g}(t,x) & =\dfrac{1}{2\pi i}\int_{\gamma-i\infty}^{\gamma+i\infty}e^{xk}dk\, C_{g}(t,k)\\
 & =\dfrac{1}{2\pi i}\int_{\gamma-i\infty}^{\gamma+i\infty}dk\,\sum_{j}\textrm{Res}\, C_{g}(s_{j}(k),k)\, e^{s_{j}(k)t}e^{xk}
\end{aligned}
\label{eq:inverseLaplace2}
\end{equation}

Finally, we transform the spatial variable to the right moving reference
frame $z=x-Vt$ where $V>0$ is the aster growth velocity we wish
to determine,

\begin{equation}
\begin{aligned}C_{g}(t,z) & =\dfrac{1}{2\pi i}\int_{\gamma-i\infty}^{\gamma+i\infty}dk\sum_{j}\textrm{Res}\, C_{g}(s_{j}(k),k)\, e^{s_{j}(k)t}e^{(z+Vt)k}\\
 & =\dfrac{1}{2\pi i}\int_{\gamma-i\infty}^{\gamma+i\infty}dk\sum_{j}\textrm{Res}\, C_{g}(s_{j}(k),k)\, e^{kz}e^{(s_{j}(k)+kV)t}
\end{aligned}
\label{eq:movingref}
\end{equation}

We must evaluate this integral in the long time limit $t\rightarrow\infty$
which corresponds to the solution that describes the front of the
expanding aster. Note that the integral takes the form of $\int f(k)e^{tg(k)}dk$
where $t$ is large. We follow 'steepest descent' or 'saddlepoint
method' and approximate the integral by the contribution from the
saddlepoint $k^{*}$ for $g_{j}(k)=s_{j}(k)-kV$. We also impose a
time invariance condition by requesting the real part of $g(k)$ to
be zero at this point. In other words, the conditions that yield the
asympotic solution are:

\begin{equation}
\left\{ \ \begin{aligned}\dfrac{d(s_{j}(k)+kV)}{dk}|_{k=k^{*}}=0\\
\textrm{Re}(s_{j}(k^{*})+k^{*}V)=0
\end{aligned}
\right.\label{eq:SteepestDescentTimeInvariance}
\end{equation}

These two equations together with the equation that specifies the
poles allow us to specify the pairs of $s_{j}$ and $k$ that describes
the shape and velocity of the expanding front (also called the 'dispersion
relations').

We return to Eq. (\ref{eq:solutionRhoGfinal}) and examine how poles
could arise. There are threee possibilities $\mathrm{det}(A)=0$,
$\mathrm{det}(A_{r})=0$, and $A_{12}^{+}+\dfrac{rA_{22}^{+}A_{r,12}}{\textrm{det}(A_{r})}=0$.
We examine them in order:

$\bullet\;$ $\mathrm{det}(A)=0\Leftrightarrow k=q_{+}$ is not a
pole, since $A_{22}\rightarrow A_{22}^{+}$, $\rho_{g}(t=0,k=q_{+},k=q_{+})=\rho_{g}^{0,+}$,
etc. and the relevant numerator becomes zero, i.e. $C_{g}(s,k)$ is
not singular and there is no residue.

$\bullet\;$ $\mathrm{det}(A_{r})=0$ does not lead to a pole either
because the second and the third term in Eq. (\ref{eq:solutionRhoGfinal})
cancel.

$\bullet\;$$A_{12}^{+}+\dfrac{rA_{22}^{+}A_{r,12}}{\textrm{det}(A_{r})}=0\Leftrightarrow\textrm{det}(A_{r})=-r(s-v_{s}q_{+}+f_{res})$
is a pole. 

Thus, the only pole of the equation Eq. (\ref{eq:solutionRhoGfinal}),
$\textrm{det}(A_{r})=-r(s-v_{s}q_{+}+f_{res})$, specifies the asymptotic,
traveling front solution of our system.

\subsection{Aster growth velocity}

The velocity of the aster expansion is derived from the conditions
imposed by Eq. (\ref{eq:SteepestDescentTimeInvariance}). It is easy
to see that both $s^{*}$ and $k^{*}$ are real numbers, so we rewrite
these conditions as follows:

\begin{equation}
V=-\dfrac{d\, s(k)}{dk}|_{k=k_{i}^{*}}=-\dfrac{s^{*}}{k^{*}}\label{eq:VeqSK}
\end{equation}

In the proceeding section, we denote $s^{*}$ as $s$ and $k^{*}$
and $k$ for simplicity. The pole is specified by $\textrm{det}(A_{r})=-r(s-v_{s}q_{+}+f_{res})\Leftrightarrow$

\begin{equation}
\begin{aligned}(s+v_{g}k+f_{cat}-r)(s-v_{s}k+f_{res})-f_{res}f_{cat}=-r(s-v_{s}q_{+}+f_{res})\quad\Leftrightarrow\\
s^{2}+[(v_{g}-v_{s})k+f_{cat}+f_{res}]s+[-v_{g}v_{s}k^{2}+(v_{g}f_{res}-v_{s}f_{cat}+v_{s}r)k-v_{s}rq_{+}]=0
\end{aligned}
\label{eq:SKnewpole}
\end{equation}

Collectively, equations (\ref{eq:quadraticForQ}), (\ref{eq:VeqSK}),
and (\ref{eq:SKnewpole}) specifies the four unknowns, $s$, $k$,
$q_{+},$ and $V$.

Since equations (\ref{eq:quadraticForQ}) and (\ref{eq:SKnewpole})
become identical when $k=q_{+}$, we reject this trivial case. Then,
these two equations leads us to

\begin{equation}
k=\dfrac{(v_{g}-v_{s})s+(v_{g}f_{res}-v_{s}f_{cat}+v_{s}r)}{v_{g}v_{s}}-q_{+}\textrm{.}\label{eq:kq}
\end{equation}

Differentiating Eq. (\ref{eq:kq}), we find

\begin{equation}
\dfrac{dk}{ds}=\dfrac{v_{g}-v_{s}}{v_{g}v_{s}}-\dfrac{dq_{+}}{ds}.\label{eq:dkds}
\end{equation}

Dividing Eq. (\ref{eq:kq}) by $s$, we find 
\begin{equation}
\dfrac{k}{s}=\dfrac{v_{g}-v_{s}}{v_{g}v_{s}}+\dfrac{v_{g}f_{res}-v_{s}f_{cat}+v_{s}r}{v_{g}v_{s}}\cdot\dfrac{1}{s}-\dfrac{q_{+}}{s}.\label{eq:kovers}
\end{equation}

Using Eq. (\ref{eq:VeqSK}), we equate equations (\ref{eq:dkds})
and (\ref{eq:kovers}) and find,

\begin{equation}
\dfrac{dq_{+}}{ds}=\dfrac{q_{+}}{s}-\dfrac{v_{g}f_{res}-v_{s}f_{cat}+v_{s}r}{v_{g}v_{s}}\cdot\dfrac{1}{s}\textrm{.}\label{eq:dqds1}
\end{equation}

Differentiating Eq. (\ref{eq:quadraticForQ}) by $s$ , we solve for
$\dfrac{dq_{+}}{ds}$ and find, 

\begin{equation}
\dfrac{dq_{+}}{ds}=\dfrac{(v_{g}-v_{s})q_{+}+2s+f_{cat}+f_{res}}{2q_{+}v_{g}v_{s}+(v_{g}-v_{s})s+v_{g}f_{res}-v_{s}f_{cat}}\textrm{.}\label{eq:dqds2}
\end{equation}

Eliminating $\dfrac{dq_{+}}{ds}$ from equations (\ref{eq:dqds1})
and (\ref{eq:dqds2}), we obtain
\[
q_{+}=\dfrac{(v_{g}f_{res}-v_{s}f_{cat}+v_{s}r)(v_{g}f_{res}-v_{s}f_{cat})+s(v_{g}^{2}f_{res}+v_{s}^{2}f_{cat}+v_{s}r(v_{g}-v_{s}))}{v_{g}v_{s}(v_{g}f_{res}-v_{s}f_{cat}+2v_{s}r)}\textrm{.}
\]

We substitute this into Eq. (\ref{eq:quadraticForQ}), and choose
the positive root for $s$.

\begin{equation}
\begin{alignedat}{1}s=\, & \dfrac{1}{(f_{cat}-r)(v_{g}+v_{s})(v_{g}f_{res}+v_{s}r)}\cdot\Big(r(v_{g}f_{res}+v_{s}f_{cat})(v_{g}f_{res}-v_{s}f_{cat}+v_{s}r)\\
 & +(v_{g}f_{res}-v_{s}f_{cat}+2v_{s}r)\sqrt{v_{g}f_{cat}f_{res}r(v_{g}f_{res}-v_{s}f_{cat}+v_{s}r)}\,\Big)
\end{alignedat}
\label{eq:sfinal}
\end{equation}

Using Eq. (\ref{eq:kq}), we find 
\begin{equation}
\begin{alignedat}{1}k=\, & \dfrac{-1}{v_{g}(f_{cat}-r)(v_{g}+v_{s})(v_{g}f_{res}+v_{s}r)}\cdot\Big(r(v_{g}f_{res}-v_{s}f_{cat}+v_{s}r)(v_{g}(f_{res}-f_{cat}+r)+v_{s}r)\\
 & +(v_{g}(f_{res}+f_{cat}-r)+v_{s}r)\sqrt{v_{g}f_{cat}f_{res}r(v_{g}f_{res}-v_{s}f_{cat}+v_{s}r)}\,\Big)
\end{alignedat}
\label{eq:kfinal}
\end{equation}

$k$ determines the rate of spatial decay of the density at the front,
which follows from Eq. (\ref{eq:movingref}).

We require $r<f_{cat}$ as $k^{*}$ diverges at $r=f_{cat}$. The
velocity of the propagating front is then given by

\begin{equation}
\boxed{\ V=-\dfrac{s}{k}=\frac{v_{g}(v_{g}f_{res}-v_{s}f_{cat})^{2}}{\left(\begin{aligned}v_{g}(v_{g}f_{res}-v_{s}f_{cat})(f_{res}+f_{cat})+(v_{g}+v_{s})(v_{g}f_{res}+v_{s}f_{cat})r\\
-2(v_{g}+v_{s})\sqrt{v_{g}f_{cat}f_{res}r(v_{g}f_{res}-v_{s}f_{cat}+v_{s}r)}
\end{aligned}
\right)}.}\label{eq:expressionV}
\end{equation}

Our expression is valid for the range $r_{c}\leq r\leq f_{cat}$.
For $r<r_{c}$, the aster fails to expand and reaches a steady state
size with limited radius. For $r>f_{cat}$, we expect some microtubules
and the microtubules they nucleated to polymerize without ever experiencing
a growth to shrinkage transition. In this scenario, we expect the
very periphery of the aster to expand at polymerization rate $v_{g}$.
We discuss $r_{c}$ in more detail in the following section.

For the special case of $J=0$, Eq. (\ref{eq:expressionV}) has the
numerator equal to zero, so we return to equations (\ref{eq:sfinal})
and (\ref{eq:kfinal}) and find the aster growth velocity as:

\begin{equation}
V=\dfrac{v_{g}(v_{g}f_{res}+v_{s}f_{cat}+2\sqrt{v_{g}v_{s}f_{cat}f_{res}})}{v_{g}(f_{res}-f_{cat}+r)+v_{s}r+\dfrac{v_{g}(f_{res}+f_{cat}-r)+v_{s}r}{v_{s}r}\sqrt{v_{g}v_{s}f_{cat}f_{res}}}\label{eq:expressionVJzero}
\end{equation}

\subsection{Critical nucleation rate and gap velocity}

We define the critical nucleation rate $r_{c}$ as the minimum value
of nucleation $r$ at which the system results in front propagation.
As seen in Eq. (\ref{eq:expressionV}), the aster expansion velocity
takes a real value as long as the term inside the square root of $v_{g}f_{res}-v_{s}f_{cat}+v_{s}r$
is positive.  For $r<r_{c}$, there is no real solution for $V$,
while, for $r>r_{c}$, a pair of solutions exists. One of them predicts
that $V$ decreases with $r$ and is therfore unphysical. Eq. (\ref{eq:expressionV})
specifies other solution of this pair.

The critical nucleation rate is:

\begin{equation}
\boxed{\quad r_{c}=f_{cat}-\dfrac{v_{g}}{v_{s}}f_{res}\quad}\label{eq:rcritical}
\end{equation}

When $r=r_{c}$, aster expansion velocity takes a finite value, which
we term the ``gap velocity''.

\begin{equation}
\boxed{V_{gap}=\underset{r\rightarrow r_{c}}{\lim}V=\dfrac{v_{g}v_{s}(-v_{g}f_{res}+v_{s}f_{cat})}{v{}_{g}^{2}f_{res}+v{}_{s}^{2}f_{cat}}}\label{eq:Vgap}
\end{equation}

Note that new microtubules nucleate only on growing plus ends; therefore,
nucleation events preferentially occur on microtubules that are in
the growing state more often than expected on average. As a result,
the subpopulation of microtubules stabilized by nucleation expands
at a velocity larger than that of a typical microtubule. In fact,
the velocity of a typical microtubule is $J$, which is negative,
while the velocity of the subpopulation of microtubules that sets
$V_{gap}$ is positive.

We also find that $V_{gap}$ is inversely proportional to the mean
microtubule length $\langle l\rangle$, 

\begin{equation}
\begin{aligned}V_{gap} & =\dfrac{v_{g}v_{s}}{v{}_{g}^{2}f_{res}+v{}_{s}^{2}f_{cat}}\cdot\frac{v_{g}f_{res}-vf_{cat}}{-v_{g}v_{s}}\cdot v_{g}v_{s}\\
 & =\dfrac{v_{g}^{2}v_{s}^{2}}{v{}_{g}^{2}f_{res}+v{}_{s}^{2}f_{cat}}\cdot\dfrac{1}{\langle l\rangle}\textrm{.}
\end{aligned}
\label{eq:VgapvsMeanlength}
\end{equation}

This points to us that the origin of $V_{gap}$ is the finite length
of microtubules in the system. The shorter the microubules are, the
more explosive the transition becomes.

In a similar manner, we can define the critical transition with respect
to any of the five parameters in the system. Thus, we expand our definition
of gap velocity to encompoass all such limits. The gap velocities
defined by the change of a single parameter are listed in Table S1.

\begin{table}
\begin{centering}
\begin{tabular}{|c|c|}
\hline 
critical parameter & $V_{gap}$\tabularnewline
\hline 
\hline 
$\begin{array}{c}
\,\\
r_{c}=f_{cat}-\dfrac{v_{g}}{v_{s}}f_{res}\\
\,
\end{array}$ & $\dfrac{v_{g}v_{s}(-v_{g}f_{res}+v_{s}f_{cat})}{v{}_{g}^{2}f_{res}+v{}_{s}^{2}f_{cat}}$\tabularnewline
\hline 
$\begin{array}{c}
\,\\
v_{g,c}=v_{s}\,\dfrac{f_{cat}-r}{f_{res}}\\
\,
\end{array}$ & $\dfrac{r(f_{cat}-r)v_{s}}{f_{cat}^{2}+f_{cat}(f_{res}-2r)+r^{2}}$\tabularnewline
\hline 
$\begin{array}{c}
\,\\
v_{s,c}=v_{g}\,\dfrac{f_{res}}{f_{cat}-r}\\
\,
\end{array}$ & $\dfrac{rf_{res}v_{g}}{f_{cat}^{2}+f_{cat}(f_{res}-2r)+r^{2}}$\tabularnewline
\hline 
$\begin{array}{c}
\,\\
f_{cat,c}=r+\dfrac{v_{g}}{v_{s}}f_{res}\\
\,
\end{array}$ & $\dfrac{rv_{g}v_{s}^{2}}{rv_{s}^{2}+f_{res}v_{g}(v_{g}+v_{s})}$\tabularnewline
\hline 
$\begin{array}{c}
\,\\
f_{res,c}=\dfrac{v_{s}}{v_{g}}(f_{cat}-r)\\
\,
\end{array}$ & $\dfrac{rv_{g}v_{s}}{-rv_{g}+f_{cat}(v_{g}+v_{s})}$\tabularnewline
\hline 
\end{tabular}
\par\end{centering}

\bigskip{}

\centering{}Table S1: Gap velocities defined by different critical
parameters.\label{table:gapvelocityy}
\end{table}

\subsection{Aster growth dynamics when $J>0$}

Past the transition to the traveling wave regime, further changes
in model parameters can make the mean polymerization rate $J$ positive.
At this point, aster growth velocity shows no unexpected behavior
and changes smoothly as $J$ changes sign (Figure 3A). The bulk state
of the aster could, however, be affected by the sign of $J$, depending
on the mode of negative feedback (see section \ref{sec:bulkdensity}
for detailed discussion). When negative feedback promotes depolymerization
at high microtubule density, $J<0$ in the bulk and asters are composed
of short microtubules that are created through nucleation and lost
through depolymerization. Thus, dynamics are essentially the same
as when $J<0$ both at the front and at the bulk. When negative feedback
simply arrests nucleation in the bulk, individual microtubules begin
to span the entire aster as in the standard model. The observations
of newly nucleated plus ends during aster growth exclude this latter
scenario \cite{Ishihara:2014ih} .

\section{Microtubule lifetime}

Consider a single microtubue nucleatied at time $t=0$. Let $\rho_{g}(t,l)$
denote the probability that it is of length $l$ at time $t$ and
in a growing state. Similarly, $\rho_{s}(t,l)$ is for the shrinking
state. Then, 

\begin{equation}
\left\{ \ \begin{aligned}\frac{\partial\rho_{g}}{\partial t}=-v_{g}\frac{\partial\rho_{g}}{\partial l}-f_{cat}\rho_{g}+f_{res}\rho_{s} & ,\\
\frac{\partial\rho_{s}}{\partial t}=+v_{s}\frac{\partial\rho_{s}}{\partial l}+f_{cat}\rho_{g}-f_{res}\rho_{s,}
\end{aligned}
\right.\label{eq:rhogrhos}
\end{equation}

with initial conditions $\rho_{g}(t=0,l)=\delta(l)$ and $\rho_{s}(t=0,l)=0$.

Assuming bounded dynamics $J<0$, we apply Laplace transforms $t\rightarrow s$
and $l\rightarrow q$, and solve the problem as before. The result
reads

\begin{equation}
\rho_{g}(s,q)=\dfrac{1}{v_{g}}\dfrac{1}{q-q_{-}(s)},\label{eq:rhogqsolution}
\end{equation}

where $q_{-}$ is the negative root of the quadratic equation (\ref{eq:quadraticForQ}).

Now, the average time spent in the growing state $\tau_{g}$, equivalent
to the rate of loss of microtubules, is given by

\begin{equation}
\begin{aligned}\tau_{g} & =\int\int\rho_{g}(t,l)dldt\\
 & =\rho_{g}(s=0,q=0)\\
 & =\dfrac{-1}{v_{g}}\left(\dfrac{v_{g}f_{res}-v_{s}f_{cat}}{v_{g}v_{s}}\right)^{-1}\\
 & =\dfrac{v_{s}}{v_{s}f_{cat}-v_{g}f_{res}}\textrm{.}
\end{aligned}
\label{eq:AvgGrowingTime}
\end{equation}

The last expression above is identical to the inverse of the critical
nucleation rate. Thus, $\tau_{g}r_{c}=1$ specifies the equation for
the critical nucleation rate, which can be intepreted as the requirement
for an average microtubule to nucleate one other microtubule during
its lifetime. 

Analogously, for the average time spent in the shrinking state $\tau_{s}$,
we find

\begin{equation}
\tau_{s}=\dfrac{v_{g}}{v_{s}f_{cat}-v_{g}f_{res}}.\label{eq:AvgShrinkingTime}
\end{equation}

Note that $\dfrac{\tau_{s}}{\tau_{g}}=\dfrac{v_{g}}{v_{s}}$. We can
also obtain total lifetime of the microtubule $\tau$ by summing over
its lifetimes in the growing and shriking states:

\begin{equation}
\tau=\tau_{g}+\tau_{s}=\dfrac{v_{g}+v_{s}}{v_{s}f_{cat}-v_{g}f_{res}}.\label{eq:MTlifetime}
\end{equation}
This result is identical to that of Bicout \cite{Bicout:1997tz} who
did not consider the lifetimes of growing and shrinking microtubules
separately.

\section{Plus end density in the aster interior\label{sec:bulkdensity}}

In the interior region of a growing aster, the density of microtubule
plus ends and microtubule length distribution are stationary and independent
of position. As a result, the increase in the number of microtubules
due to nucleation must equal microtubule loss. For growing microtubules,
the rate of loss is given by $\dfrac{1}{\tau_{g}}$ (see Eq. (\ref{eq:AvgGrowingTime}))
while the rate of gain is simply the nucleation rate. Thus,

\begin{equation}
r^{bulk}=\dfrac{1}{\tau_{g}^{bulk}}.\label{eq:rbulktaubulk}
\end{equation}

\subsection{Nucleation changes with plus end density}

Logistic function is a commonly used mechanism for negative feedback
in the context of expanding populations (see \cite{Korolev:2013bc}
for an example). In Eq. (\ref{eq:nucleationbifurcation}), it takes
the following form

\begin{equation}
r=r_{0}\left(1-\dfrac{C_{g}}{K}\right),\label{eq:logisiticform}
\end{equation}

where $r_{0}$ is the nucleation rate at low plus end densities, and
$K$ sets the scale of $C_{g}$ when the negative feedback becomes
appreciable, resulting in stationary plus end density. The balance
between microtubule production and loss given by Eq. (\ref{eq:rbulktaubulk})
results in the following expression for the plus end density in the
bulk

\begin{equation}
C_{g}^{b}=K\left(\dfrac{r-r_{c}}{r}\right)\textrm{,}\label{eq:SSdensityLogisitic}
\end{equation}

where we used the fact that $\dfrac{1}{\tau_{g}}=r_{c}$.

Michaelis-Menten type kinetics is an alternative functional form for
the negative feedback that could arise, for example, due to the limitation
of a nucleating factor,

\begin{equation}
r=\dfrac{r_{0}}{1+\dfrac{C_{g}}{K}}\textrm{.}\label{eq:MMform}
\end{equation}
This results in the following plus end density:

\begin{equation}
C_{g}^{b}=K\left(\dfrac{r-r_{c}}{r_{c}}\right)\textrm{.}\label{eq:SSdensityMM}
\end{equation}

In either case, $C_{g}^{b}$ is proportional to $r-r_{c}$ close to
the transition. Fluctuations in $C_{g}^{b}$ due to the stochasticity
of microtubule nucleation and collapse can alter this behavior to
$(r-r_{c})^{\beta}$, where $\beta$ is the corresponding exponent
of a non-equilibrium transition.%
\footnote{This transition most likely belongs to the directed percolation universality
class \cite{Hinrichsen:2000cy}.%
} Note that the critical nucleation rate used here is the same as in
Eq. (\ref{eq:rcritical}). In particular, all the values of all the
model parameters are obtained in the limit of small $C_{g}$, i.e.
at the edge of the aster.

\subsection{Catastrophe rate changes with plus end density\label{sub:CatChange}}

Instead of changing the nucleation rate, the cell can promote microtubule
depolymerization to ensure that the bulk density does not grow indefinitely.
Consider the negative feedback such that the nucleation is constant
throughout the aster

Specifically, the nucleation term is simply proportional to the local
density of growing plus ends as in

\begin{equation}
Q(t,x)=r\cdot C_{g}(t,x)\textrm{,}\label{eq:constantnuc}
\end{equation}

while we implement feedback regulation in the catastrophe rate,

\begin{equation}
f_{cat}=f_{cat}^{0}\left(1+\dfrac{C_{g}}{K}\right)\textrm{.}\label{eq:RegFcat}
\end{equation}

Here, $f_{cat}^{0}$ corresponds to the catastrophe rate at the leading
edge of the growing aster where $C_{g}$ is small and $K$ specifies
the plus end densities at which the negative feedback becomes appreciable.

The balance between the constant nucleation rate and the loss rate
that increases with the plus end densities leads to the following
solution for the steady-state density in the bulk:

\begin{equation}
C_{g}^{b}=K\left(\dfrac{r-r_{c}}{f_{cat}^{0}}\right)\textrm{.}\label{eq:SSdensityRegFcat}
\end{equation}
Note that the critical nucleation rate used here is the same as in
Eq. (\ref{eq:rcritical}). In particular, all the values of all the
model parameters are obtained in the limit of small $C_{g}$, i.e.
at the edge of the aster.

\subsection{Depolymerization rate changes with plus end density}

As a final example, we consider a situation where the nucleation is
constant as in Eq. (\ref{eq:constantnuc}), while the depolymerization
rate increases with plus end density,

\begin{equation}
v_{s}=v_{s}^{0}\left(1+\dfrac{C_{g}}{K}\right)\textrm{.}\label{eq:RegVs}
\end{equation}

Here, $v_{s}^{0}$ corresponds to the depolymerization rate at the
leading edge of the growing aster where $C_{g}$ is small. For this
feedback mechanism, we find

\begin{equation}
C_{g}^{b}=K\left(\dfrac{r-r_{c}}{f_{cat}-r}\right)\textrm{.}\label{eq:SSdensityRegDepoly}
\end{equation}

In the above four examples we found that close to the onset of aster
growth the bulk density is proportional to $r-r_{c}$, and it is easy
to see that this is true regardless of the feedback mechanism. Indeed,
at the critical transition, the nucleation barely keeps up with loss
at the front; thus, an infinitesimal increase in density and the corresponding
negative feedback would alter the balance and $C_{g}$ must be zero
at $r=r_{c}$. As a result,  $C_{g}^{b}$ is proportional to $r-r_{c}$
just above to the transition. In contrast, the expansion velocity
near the transition does not vanish and remains at a high value specified
by $V_{gap}$. In consequence, the cell can control the density of
the microtubules in the aster and, therefore, its mechanical properties
by small changes in the nucleation rate without significantly altering
the kinetics of aster growth.

\section{Other types of feedback regulation lead to the same explosive transition\label{sub:Other-types-of}}

Apart from the carrying capacity on nucleation kinetics considered
above, other forms of negative feedback on the system are possible.
This may include scenarios such as decreasing polymerization rate
or increasing catastrophe rate with higher local density of microtubules.
Feedback regulation at higher microtubule densities is important in
the interior of the growing aster while the linearized equations solved
above capture the dynamics of the leading edge. Thus, different forms
of feedback regulation lead to the same critical transition predicted
by equations (\ref{eq:rcritical}) and (\ref{eq:expressionV}).

Consider one of such alternate scenario for (\ref{eq:PDEsystem})
as described in section \ref{sub:CatChange}. Here, the negative feedback
is implemented at the level of catastrophe rate instead of nucleation.
We numerically solved the partial differential equations under these
assumptions. Similar to the previous case, we observed the emergence
of propagating fronts in a parameter dependent manner (Figure 3-figure
supplement 1A). Although the shape of the propogating front is different,
the aster growth velocity is again in excellent agreement with our
analytical solution (Figure 3-figure supplement 1B).

\section{Estimation of unknown parameters $f_{cat}$ and $r$}

By combining analytical solutions and experimental measurements in
frog egg extract, we estimate the values of unknown parameters in
our model. Given our direct measurements for $v_{g}$, $v_{s}$, $f_{cat}$,
$V$, and $V_{gap}$ , we have two unknowns $f_{cat}$ and $r$. To
simultaneously estimate these values, we need two equations.

The first is the aster growth velocity equation (\ref{eq:expressionV}).
For the second equation, we use one of the equations for $V_{gap}$
as shown in Table S1, which correspond to different assumptions on
how $V_{gap}$ was achieved by the MCAK-Q710 perturbation. The result
of the parameter estimations are summarized in Table S2. In all scenarios,
the values of $f_{res}$ and $r$ are in relative agreement.

\begin{table}[h]
\begin{centering}
\begin{tabular}{ccccc}
estimated parameter & units & $r\rightarrow r_{c}$ & $f_{res}\rightarrow f_{res,c}$ & $v_{s}\rightarrow v_{s,c}$\tabularnewline
\hline 
\hline 
$f_{res}$ & $min^{-1}$ & 2.0$\pm$0.3 & 3.0$\pm$0.7 & 3.0$\pm$0.7\tabularnewline
$r$ & $min^{-1}$ & 2.1$\pm$0.2 & 1.9$\pm$0.2 & 1.8$\pm$0.2\tabularnewline
$K$ & $\mu m^{-1}$ & 0.053$\pm$0.030 & 0.12$\pm$0.09 & 0.15$\pm$0.10\tabularnewline
$\langle l\rangle$ & $\mu m$ & 16$\pm$2 & 32$\pm$34 & 39$\pm$44\tabularnewline
\hline 
\end{tabular}
\par\end{centering}

\bigskip{}

\raggedright{}Table S2: Estimated parameter values for different scenarios
on how MCAK-Q710 arrested aster growth. Different expressions for
$V_{gap}$ shown in Table S1 were used. In all cases, the values of
$v_{g}$, $v_{s}$, $f_{cat}$, $V$, and $V_{gap}$ were the same
as in Table 1. \label{table:fittingalternate}
\end{table}

\section{Aster growth by polymer-stimulated nucleation of microtubules}

In this section, we consider a scenario where microtubule nucleation
is stimulated by the local density of polymer rather than the density
of growing plus ends. Although we have not obtained the analytical
solution for this scenario, we derive the expression for the critical
nucleation rate for aster growth and confirm these results by numerical
simulations. Importantly, the transition from stationary to growing
asters predicts a finite jump in velocity.

\subsection{Critical nucleation rate for polymer-stimulated nucleation}

Let $p$ denote the polymer-stimulated nucleation rate with units
$[{\rm time}^{-1}\,{\rm microtubule\, length}^{-1}]$. A microtubule
of length $l$ with nucleate $p\, l\, dt$ microtubules in time $dt$.

To derive the critical nucleation rate $p_{c}$ for aster growth,
we require that a single microtubule during its entire liftime must
nucleate at least one microtubule:

\begin{equation}
p_{c}\int_{0}^{\infty}dt\int_{0}^{\infty}dl\, l\,\rho(t,l)=1,\label{eq:condpolymerbasednucleation}
\end{equation}

where $\rho(t,l)$ denotes the local density of all plus ends. As
plus ends are either in the growing or shrinking states, $\rho(t,l)=\rho_{g}+\rho_{s}$.

By applying the Laplace transforms $t\rightarrow s$ and $l\rightarrow q$
to the left hand side of Eq. (\ref{eq:condpolymerbasednucleation}),
we obtain

\begin{equation}
\int_{0}^{\infty}dt\int_{0}^{\infty}dl\, l\,\rho(t,l)=\int_{0}^{\infty}\rho(s=0,l)\, l\, dl=-\dfrac{d\rho(s=0,q)}{dq}|_{q=0}.\label{eq:Laplacecondpolymer}
\end{equation}

Thus, we may solve for the desired critical nucleation rate as $p_{c}=\left(-\dfrac{d\rho(s=0,q)}{dq}|_{q=0}\right)^{-1}$.

We have previously obtained the expression for $\rho_{g}$ as Eq.
(\ref{eq:rhogqsolution}). To obtain the equivalent expression for
$\rho_{s}$, we return to the dynamic equation Eq. (\ref{eq:rhogrhos})
and apply the Laplace transforms $t\rightarrow s$ and $l\rightarrow q$,

\begin{equation}
s\rho_{g}-1=-v_{g}q\rho_{g}-f_{cat}\rho_{g}+f_{res}\rho_{s},\label{eq:Laplaceofdynamiceqn}
\end{equation}

which yields,

\begin{equation}
\rho_{s}(s,q)=\dfrac{\rho_{g}(s+v_{g}q+f_{cat})-1}{f_{res}}.\label{eq:rhosqsolution}
\end{equation}

Combining Eq. (\ref{eq:rhogqsolution}) and (\ref{eq:rhosqsolution}),
we obtain

\begin{equation}
\rho(s,q)=\rho_{g}+\rho_{s}=-\dfrac{1}{f_{res}}+\dfrac{s+v_{g}q+f_{cat}}{v_{g}f_{res}}\cdot\dfrac{1}{q-q_{-}}+\dfrac{1}{v_{g}}\cdot\dfrac{1}{q-q_{-}}.\label{eq:rhogsqsolution}
\end{equation}

Differentiating and setting $q=0$, we find

\begin{equation}
\begin{alignedat}{2}-\dfrac{d\rho(s=0,q)}{dq}|_{q=0} & = & \dfrac{1}{v_{g}f_{res}}\cdot\dfrac{v_{g}q_{-}+f_{cat}}{q_{-}^{2}}+\dfrac{1}{v_{g}}\cdot\dfrac{1}{q_{-}^{2}}\\
 & = & \dfrac{1}{v_{g}q_{-}^{2}}\left(1+\dfrac{f_{cat}}{f_{res}}\right)+\dfrac{1}{f_{res}q_{-}}.
\end{alignedat}
\label{eq:drhodq}
\end{equation}

From Eq. (\ref{eq:quadraticForQ}), we find

\begin{equation}
q_{-}(s=0)=-\dfrac{v_{s}f_{cat}-v_{g}f_{res}}{v_{g}v_{s}}.\label{eq:qminusS0}
\end{equation}

Using Eq. (\ref{eq:drhodq}) and (\ref{eq:qminusS0}), we solve for
the critical nucleation rate

\begin{align}
p_{c} & =\left(-\dfrac{d\rho(s=0,q)}{dq}|_{q=0}\right)^{-1}\nonumber \\
 & =\dfrac{f_{res}v_{g}q_{-}^{2}}{f_{res}\left(1+\dfrac{f_{cat}}{f_{res}}\right)+v_{g}q_{-}}\nonumber \\
 & =f_{res}v_{g}\cdot\dfrac{(v_{s}f_{cat}-v_{g}f_{res})^{2}}{v_{g}^{2}v_{s}^{2}}\cdot\dfrac{1}{f_{res}+f_{cat}-f_{cat}+\dfrac{v_{g}}{v_{s}}f_{res}}\nonumber \\
 & =\dfrac{(v_{s}f_{cat}-v_{g}f_{res})^{2}}{v_{g}v_{s}(v_{g}+v_{s})}.\label{eq:pccalc}
\end{align}

In the scenario of polymer-stimulated nucleation, the minimal nucleation
rate required for aster growth is

\begin{equation}
\boxed{\quad p_{c}=\dfrac{(v_{s}f_{cat}-v_{g}f_{res})^{2}}{v_{g}v_{s}(v_{g}+v_{s})}\quad}.\label{eq:pcritical}
\end{equation}

When both types of nucleation are present, we expect $\dfrac{r}{r_{c}}+\dfrac{p}{p_{c}}=1$
to define the transition, where $r_{c}$ and $p_{c}$ are defined
in the absence of the other type of nucleation as in Eq. (\ref{eq:rcritical})
and (\ref{eq:pcritical}).

\subsection{Numerical simulations predict a gap velocity for aster growth by
polymer-stimulated nucleation}

We modified our numerical simulation to ask if polymer-stimulated
nucleation predicts the aster growth. Similar to the scenario of growing-plus-end-stimulated
nucleation (Fig. 2A), low nucleation rate predicts a stationary aster
(Figure 3-figure supplement 2A, left), while high nucleation rate
predicts an aster that continuously increases in radius (Figure 3-figure
supplement 2A, right) even when individual microtubules are unstable
($J<0$). To systematically explore the polymer-stimulated nucleation
scenario, we varied the model parameters and measured the aster growth
velocity $V$. We find that the transition from a stationary to a
growing aster is accompanied by a finite jump in $V$ (Figure 3-figure
supplement 2B and 2C). Our predicitions for the critical polymer nucleation
rates $p_{c}$ is in excellent quantitative agreement.

\subsection{Comparison of autocatalytic nucleation mechanisms and predictions
for aster growth}

Here, we compare and summarize the theoretical predictions of growing-plus-end-stimulated
nucleation vs. polymer-stimulated nucleation. Both scenarios predict
\begin{enumerate}
\item stationary and continuously growing asters in a parameter dependent
manner
\item the feasibility of aster growth with $J<0$, and that such asters
are composed of short microtubules
\item explosive transition to growth, or ``gap velocity'', which allows
independent control of aster density and growth velocity.
\end{enumerate}
The two scenarios predict qualitatively different transitions when
the the nucleation rate is increased and the aster growth velocity
reaches $V=v_{g}$. With growing-plus-end stimulated nucleation, $V$
approached $v_{g}$ in a smooth manner (Fig. 3). In contrast, the
polymer-stimulated nucleation predicted a finte jump of $V$ to $v_{g}$
(Figure 3-figure supplement 2A, right). In the future, it may be possible
to exploit this difference to distinguish the two scenarios of nucleation
experimentally.

\section{Gap velocity constrains possible models of aster growth}

Here, we describe three examples of potential aster growth models
that are readily rejected by the existence of a gap velocity. In all
cases, let us assume that microtubules are unstable (bounded dynamics
$J<0$) with finite lifetime. Note that all three models do not account
for the internal dynamics of agents, namely, the length information
of individual microtubules.

\subsection{A simple expanding shell model with autocatalytic}

Consider an aster growth model that does not keep track of microtubule
positions, but simply translates the number of microtubules into aster
size:

\begin{equation}
\begin{array}{c}
\dfrac{dN_{+}}{dt}\sim\left(r-\dfrac{1}{\tau}\right)N_{+}^{\alpha}\\
R^{d}\sim N_{+}
\end{array}\label{eq:shellmodel}
\end{equation}

$N_{+}$ is the number of plus ends, $R$ is the aster radius, $d$
is the number of spatial dimensions, $r$ is the nucleation rate,
and $\tau$ is the microtubule lifetime. $\alpha=1$ corresponds to
no negative feedback, while $\alpha=\dfrac{d-1}{d}$ corresponds to
nucleation only at aster periphery. This is the simplest, virtually
no spatial model. For $\alpha=1$, the growth is exponential in time
rather than linear. For $\alpha=\dfrac{d-1}{d}$, the growth is linear
with the velocity $V\sim\left(r-\dfrac{1}{\tau}\right)$. Thus, the
model exhibits critical nucleation (i.e. both stationary and growing
asters), but no gap velocity.

\subsection{A reaction-diffusion model}

Previously, we hypothesized a Fisher-Kolmogorov type, reaction-diffusion
model of aster growth focusing on plus end dynamics \cite{Ishihara:2014gpa}
and autocatalytic nucleation. Denoting the plus end density as $C_{+}$,
carrying capacity as $K$, and the effective growth rate as $r-\dfrac{1}{\tau}$,
the model is as follows:

\begin{equation}
\dfrac{\partial C_{+}}{\partial t}=D\dfrac{\partial^{2}C_{+}}{\partial x^{2}}+\left(r-\dfrac{1}{\tau}\right)\, C_{+}\left(1-\dfrac{C_{+}}{K}\right)\label{eq:fishersimple}
\end{equation}

This predicts an aster growth velocity of $V\sim\sqrt{D\left(r-\dfrac{1}{\tau}\right)}$
for $r>\dfrac{1}{\tau}$ with no gap velocity.

\subsection{A reaction-diffusion model with cooperative nucleation}

A more general reaction-diffusion model is obtained by replacing the
logistic growth term in Eq. (\ref{eq:fishersimple}) by an arbitrary
nonlinear function of the plus end density $F(C_{+})$:

\begin{equation}
\dfrac{\partial C_{+}}{\partial t}=D\dfrac{\partial^{2}C_{+}}{\partial x^{2}}+F(C_{+})\label{eq:fishergeneral}
\end{equation}

$F(C_{+})$ can specify whether there is a minimal concentration of
microtubules necessary for nucleation or account for other effects
such as cooperativity. Despite the possibilities of quite complicated
nucleation dynamics, all reaction-diffusion models specified by Eq.
(\ref{eq:fishergeneral}) exhibit no gap velocity \cite{Murray:2002uo}.
Gap velocity has also not been observed in a variety of further extensions
of Eq. (\ref{eq:fishergeneral}) that account for advection terms
and density-dependent diffusion \cite{Murray:2002uo}.

\pagebreak{}\includegraphics[scale=0.7]{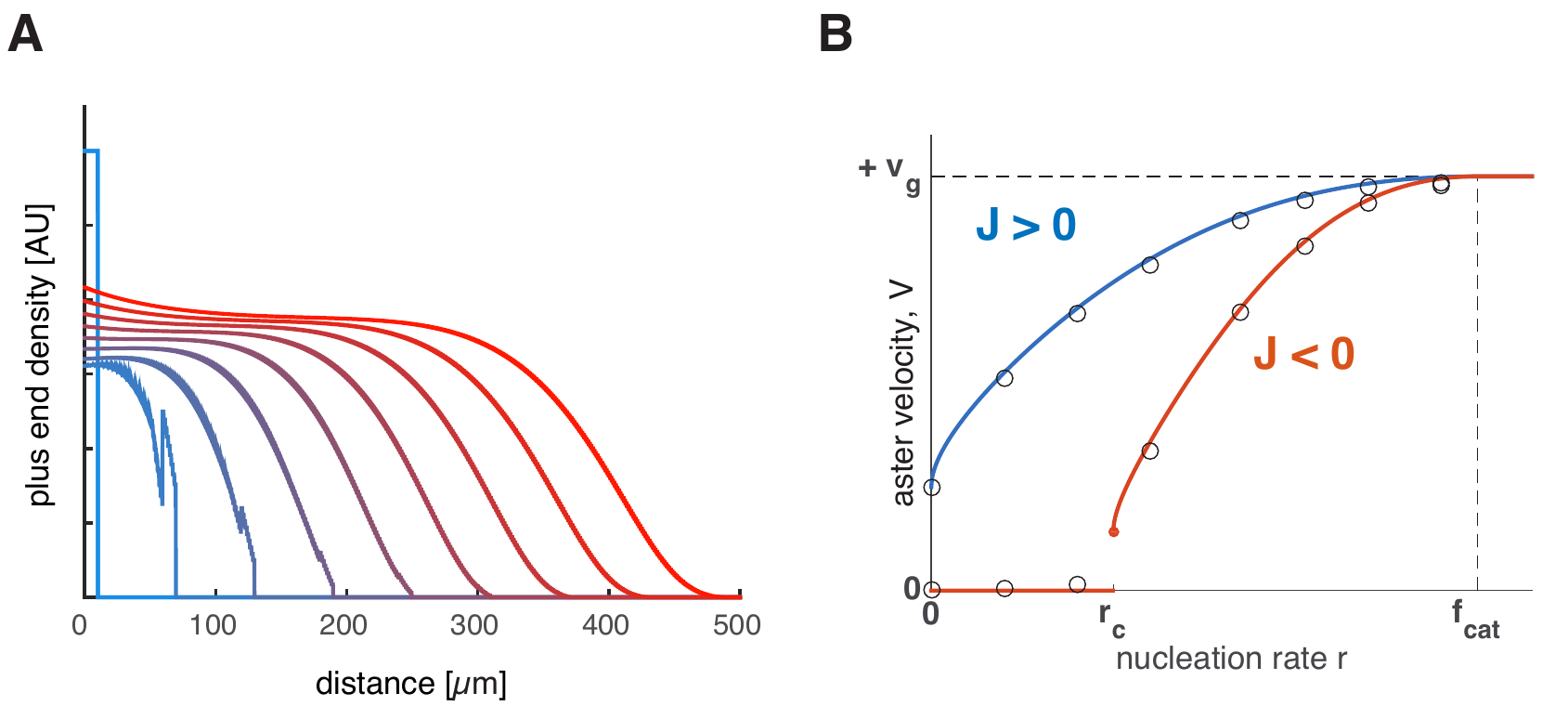}

\subsubsection*{Figure 3-figure supplement 1}

\begin{singlespace}
\textbf{Feedback regulation of catastrophe rate lead to the same explosive
transition.} (A) Time evolution of growing plus end density similar
to Fig. 2. This example represent a scenario where the nucleation
rate is above the critical nucleation rate ($v_{g}=30,v_{s}=40,f_{cat}=3,f_{res}=1,r=2.5$)
resulting in aster growth. (B) Analytical solution (lines) and numerical
simulations (dots) predict aster growth velocity as a function of
nucleation rate similar to Fig. 3A. Blue line corresponds to $J>0\,(v_{g}=30,v_{s}=15,f_{cat}=3,f_{res}=3)$
and red line to $J<0\,(v_{g}=30,v_{s}=15,f_{cat}=3,f_{res}=1)$.
\end{singlespace}

\pagebreak{}

\subsubsection*{\protect\includegraphics[scale=0.7]{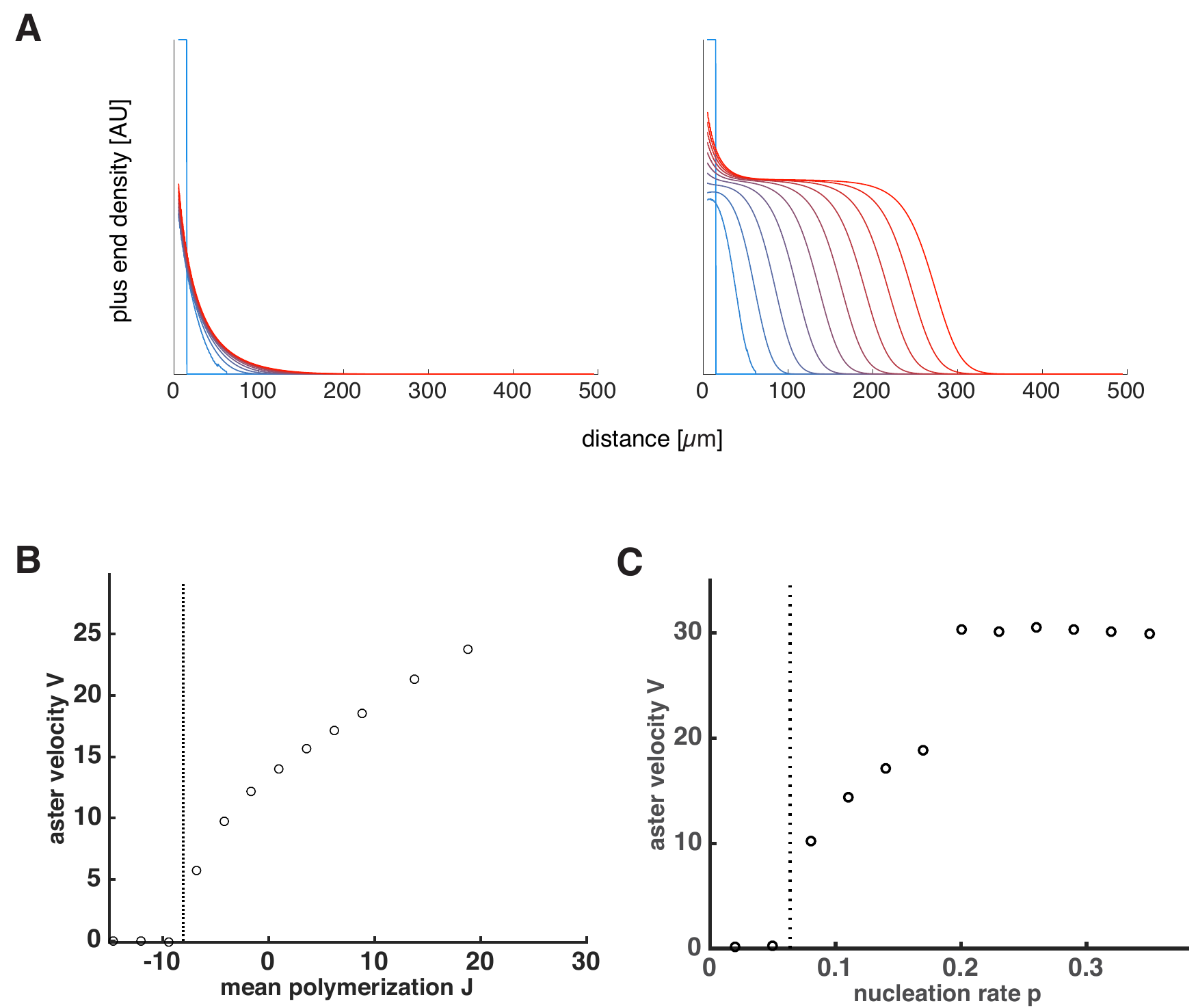}}

\subsubsection*{Figure 3-figure supplement 2}

\begin{singlespace}
\textbf{Aster growth by polymer-stimulated nucleation lead to the
same explosive transition.} (A) Time evolution of growing plus end
density similar to Fig. 2A. Below the critical nucleation rate, asters
are stationary (left, $v_{g}=30,v_{s}=40,f_{cat}=3,f_{res}=1,p=0.07$).
Above the critical nucleation rate, asters grow in radius (right,
$p=0.18$) even when microtubules are unstable ($J<0$). Here, the
critical polymer nucleation rate $p_{c}=0.0964...$ as predicted by
Eq. (\ref{eq:pcritical}). (B) Numerical simulations predict aster
growth velocity as a function of $J$ ($f_{res}$ was varied while
keeping $v_{g}=30,v_{s}=15,f_{cat}=3,p=0.04$). (C) Numerical simulations
predict aster growth velocity as a function of polymer-stimulated
nucleation rate $p$ ($p$ was varied while keeping $v_{g}=30,v_{s}=15,f_{cat}=3,f_{res}=0.3$).
Dashed vertical lines indicate the predicted critical transitions
from Eq. (\ref{eq:pcritical}).
\end{singlespace}

\pagebreak{}

\subsubsection*{\protect\includegraphics[scale=0.7]{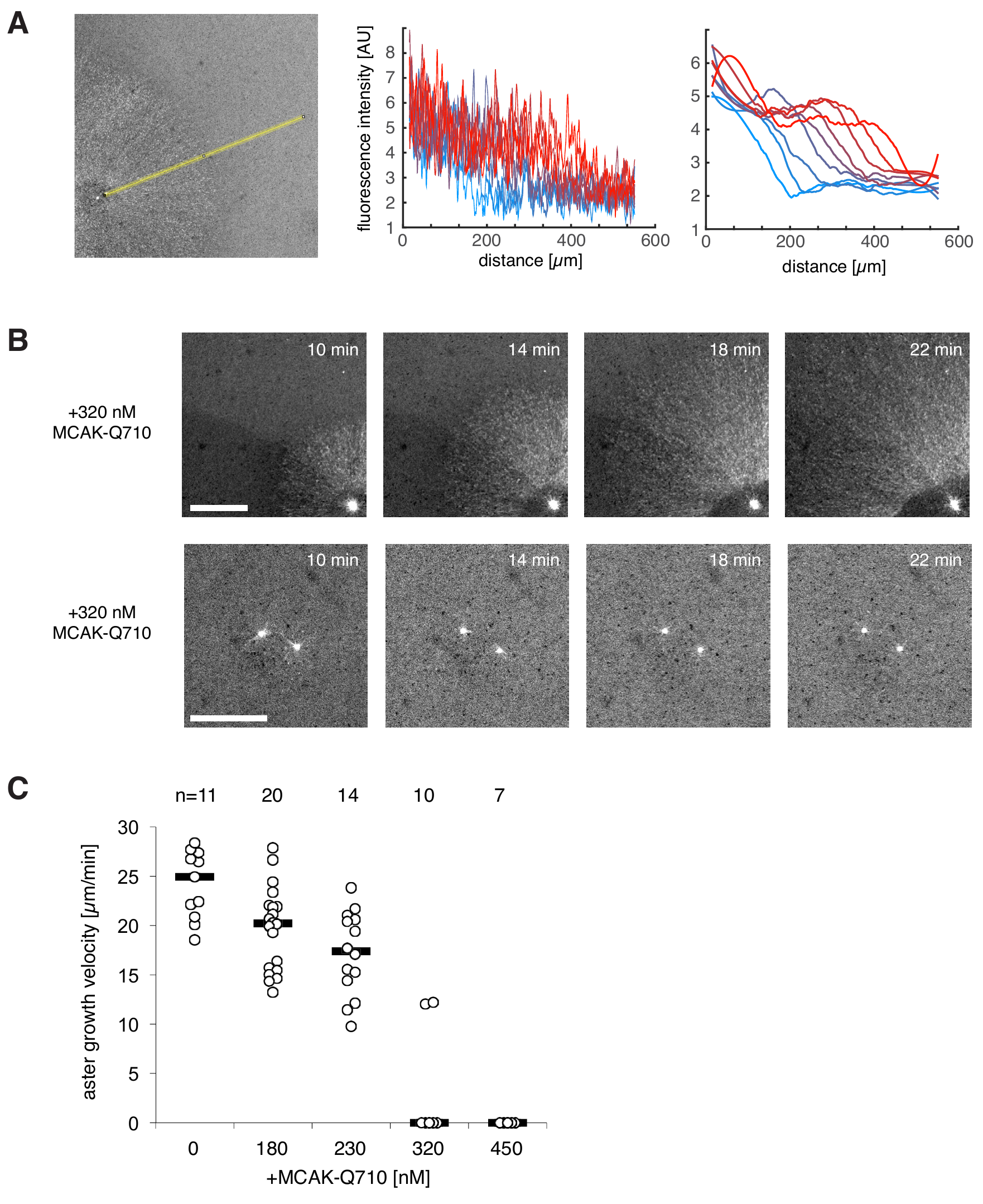}}

\pagebreak{}

\subsubsection*{Figure 4-figure supplement 1}

\begin{singlespace}
\textbf{Aurora A kinase bead asters at different MCAK-Q710 concentrations.}
(A) Measuring aster growth velocities from time-lapse images of asters
visualized with the plus end marker EB1-mApple. A linear region is
chosen in the radial outward direction (left). The raw fluorescent
intensity profiles (center) are subjected to a low pass filter (right),
and the half-max position was manually selected to define the radius
of the aster at different time points. Blue to red lines indicate
profiles at two minute intervals. (B) At the critical concentration
of 320 nM MCAK-Q710, some asters assembled from Aurora A beads showed
slow growth (top) while others contained few microtubules which gradually
decreased over time (bottom). The latter were scored as zero growth
velocity. The reaction was started at time zero by the addition of
calcium and beads to the extract. Scale bars 100 $\mu$m. (C) Aster
growth velocities measured at increasing MCAK-Q710 concentrations.
Biological replicate of the same experiment as in Fig. 4B. 
\end{singlespace}

\pagebreak{}

\subsubsection*{\protect\includegraphics[scale=0.7]{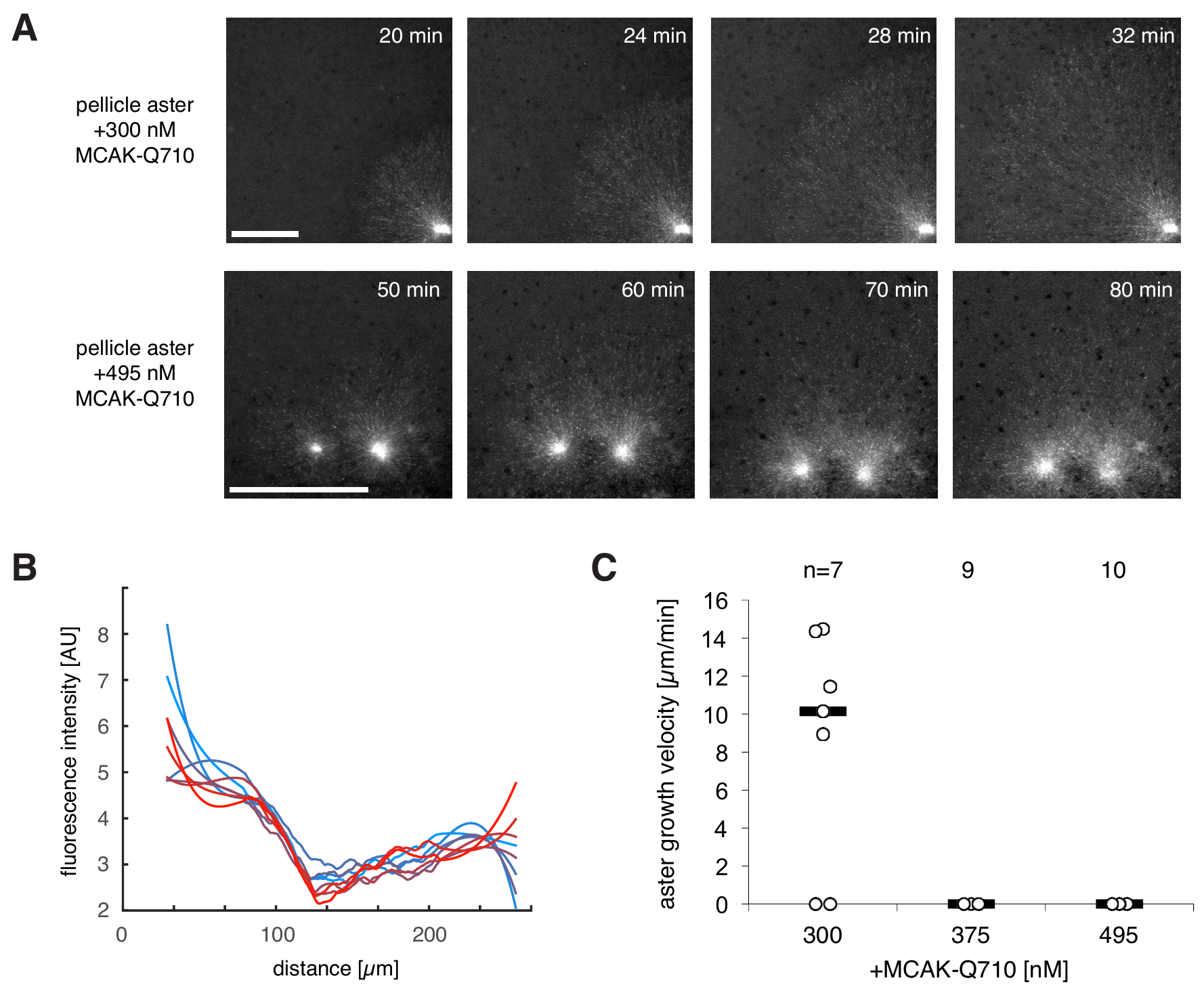}}

\subsubsection*{Figure 4-figure supplement 2}

\begin{singlespace}
\textbf{Pellicle asters at different MCAK-Q710 concentrations.} (A)
Asters assembled by \textit{Tetrahymena} pellicles as the nucleating
center showed aster growth which was slowed down by MCAK-Q710 (top).
At higher MCAK-Q710 concentrations, stationary asters that did not
change its radius for over 60 min were observed (bottom). (B) EB1-mApple
fluorescence intensity profile of the stationary aster in panel A
for time intervals 70-84 min post calcium addition. Such asters were
scored as zero velocity. (C) Pellicle aster growth velocities at different
MCAK-Q710 concentrations. 
\end{singlespace}

\pagebreak{}

\includegraphics[scale=0.7]{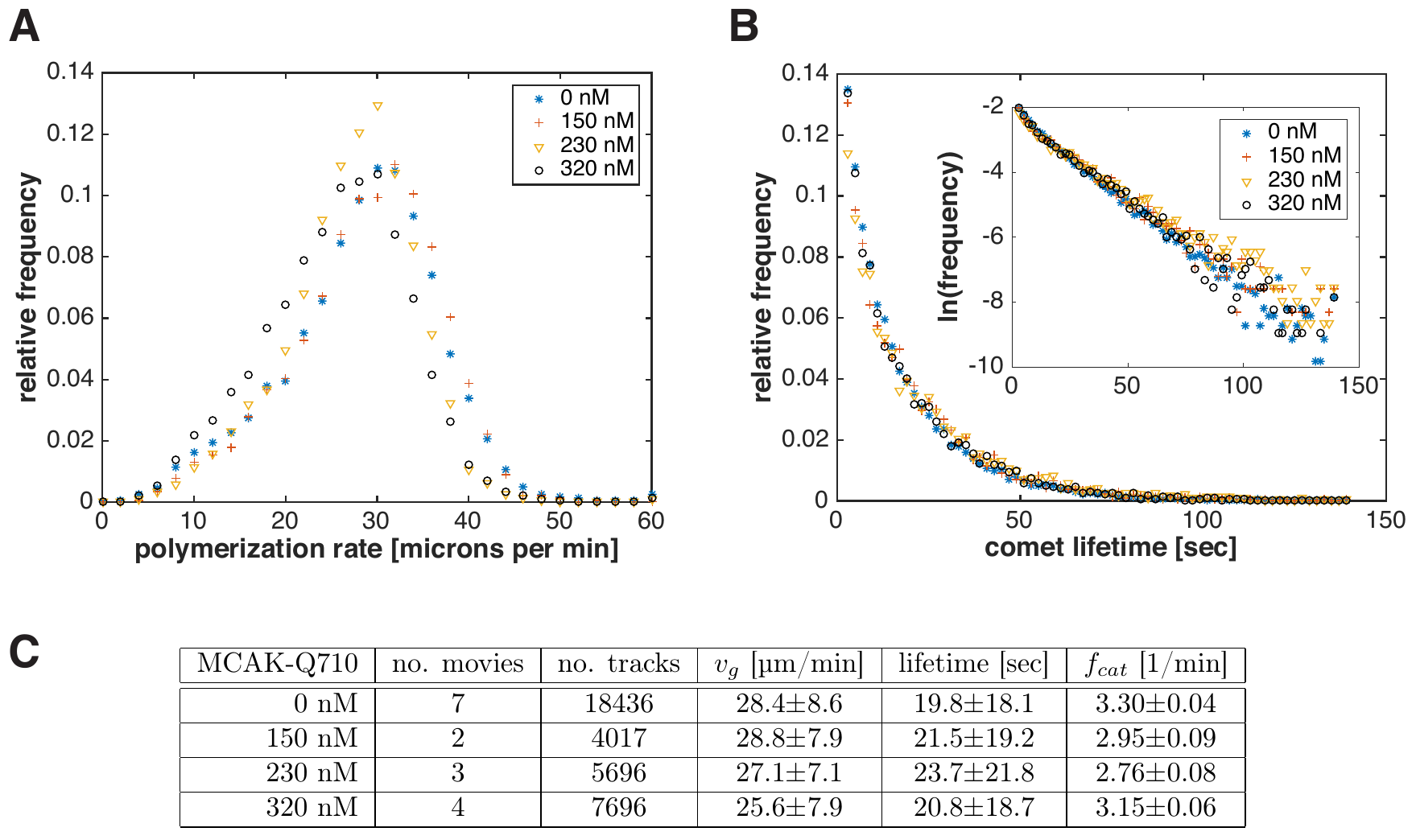}

\subsubsection*{Figure 4-figure supplement 3}

\begin{singlespace}
\textbf{Plus end polymerization rate and catastrophe rate do not significantly
change with MCAK-Q710 titration.} Measurements were made by imaging
and tracking EB1 comets in growing interphase asters assembled by
Aurora A beads (see Methods). (A) Distribution of plus end polymerization
rates at different MCAK-Q710 concentrations. (B) Distribution of EB1
comet lifetimes at different MCAK-Q710 concentrations. Inset shows
the same data plotted on a semilog scale. (C) Summary of measurements
from EB1 tracking analysis. The table shows the number of movies (or
asters) and total number of tracks analyzed for each condition. Errors
indicate standard error. The catastrophe rate $f_{cat}$ was derived
from a linear fit to the semilog plots of the lifetime distributions
in the intervals 5-60 seconds. Its mean and standard error were calculated
by bootstrapping.\end{singlespace}


\clearpage
\singlespacing

\bibliography{mybib}
\bibliographystyle{apalike}

\end{document}